\documentclass[11pt,a4paper]{article}
\usepackage[margin=2.5cm]{geometry}
\usepackage{graphicx}
\usepackage{booktabs}
\usepackage{amsmath}
\usepackage{hyperref}
\usepackage{natbib}
\usepackage{caption}
\usepackage{float}

\title{The Little Scientist: What Does the Scientific Method Buy an LLM Agent?}
\author{Travis Smith\\
\small Independent Researcher, S\~ao Miguel, Azores, Portugal\\
\texttt{travis.smith@fiveminutesforward.com}\\
\small ORCID: 0009-0009-6732-445X}
\date{First draft August 2026; updated November 3, 2026}

\begin{document}
\maketitle

\begin{abstract}
\noindent What happens when you teach an LLM-based agent the scientific method?

\noindent\textbf{Motivation:} Scientific discovery emerges from cycles of hypothesis, implementation, empirical testing, and feedback. Can this process be automated? We approach automated algorithm design through the lens of the scientific method, where an LLM-based agent goes through each step of the process in an ordered, iterative fashion.

\noindent\textbf{Results:} We present \emph{The Little Scientist}, a framework that maps the scientific method onto an LLM-based agent in a loop: a Scientist agent iterates hypothesis, prediction, code, test, and reconciliation inside an evaluation environment that returns structured diagnostics, and a Kuhn agent injects paradigm-shifting reframings at plateaus. On DNA motif discovery, the process wrote an algorithm from scratch (\emph{DALE}) that outperforms STREME across 132 ENCODE transcription factors (TFs) on the shuffled-negative benchmark (mean AUROC 0.842 vs.\ 0.803) while running $11\times$ faster. We then asked \emph{why} this method worked, and that question became the paper's scientific core: a preregistered, controlled teardown of the framework's own scaffolding consisting of ten ablation arms against one-shot bandit nulls. The answer overturned the premise. Seven of ten loop arms fail to beat a bandit drawing samples from the same model and prompt, the full scaffold performs below blind sampling at $7{-}8\times$ the token cost, and a floor-prompt arm with no framework at all wins on both score and value-per-token. On a held-out genomic benchmark, the best loop artifact and the best bandit draw agree to within 0.002, and the DALE algorithm from the initial experiment scored similarly. The convergent picture is that LLM agents do not reason their way past what training placed in the weights. Auto-research, on this evidence, is sampling, selection, and remixing, which is the bet the Evolve-family architectures made when they chose populations and external selection over scratchpad reasoning, and why those choices were right.

\end{abstract}

\section{Introduction}
\label{sec:intro}

Agent-driven algorithm discovery has produced impressive results. AlphaEvolve \citep{novikov2025alphaevolve} and FunSearch \citep{romera2024funsearch} discovered algorithms competitive with human-designed ones using LLM-based evolutionary search. The Evolution of Heuristics framework \citep{liu2024eoh} applied similar ideas to automatic heuristic design. \citet{li2026agentic} propose teams of LLM agents as computational research teams. These systems share a common approach: evolutionary or genetic operators on code, guided by a scalar fitness signal.

Recent theoretical analysis argues that this generate-and-rank paradigm misses the problem of sparse feedback, and that auto-research should instead use dense intermediate signals---analogous to coverage-guided fuzzing---to direct search rather than merely ranking completed samples~\citep{he2026agentic}.

Human scientists do not work this way. They form hypotheses before experimenting, predict outcomes, reconcile predictions with results, and when progress stalls, step back and reframe the problem entirely --- a process Thomas Kuhn \citep{kuhn1962structure} described as the alternation between normal and revolutionary science. We ask: can an LLM-based agent that follows the scientific method --- rather than evolutionary search --- discover competitive algorithms?

We present \emph{The Little Scientist}, a framework built around a single organizing idea: an LLM-based agent that behaves like a scientist. The Scientist agent works inside an evaluation environment that benchmarks its code against real data and feeds results back with structured diagnostics. The LLM agent walks through the scientific method --- hypothesis, prediction, code, test, reconciliation --- for each iteration. When the Scientist plateaus at a local optimum, a second agent (the Kuhn agent, named after Thomas Kuhn's \citep{kuhn1962structure} distinction between normal and revolutionary science) injects a paradigm-shifting conjecture: a domain-specific reframing paired with an inspiration drawn from a different scientific discipline. This forces the Scientist to explore a different region of the LLM's latent space, potentially breaking out of the local optimum.

We demonstrate the framework on a problem that requires genuine de novo algorithm design:

\begin{itemize}
    \item \textbf{DNA motif discovery} (de novo algorithm design). Given ChIP-seq peak sequences, discover position weight matrices (PWMs) representing TF binding specificity. The Scientist wrote an algorithm from scratch --- \emph{DALE} (\textbf{D}ual-seed \textbf{A}lgorithm for \textbf{L}atent \textbf{E}numeration) --- that significantly outperforms STREME \citep{bailey2021streme} (the default in the MEME Suite) across 132 ENCODE TFs, while running 11$\times$ faster. This demonstrates a different capability: the framework can produce genuinely novel algorithms, not just optimize combinations of existing tools.
\end{itemize}

The algorithm is the unmodified output of the autonomous search process.

A notable property of this approach is that the outputs are \emph{interpretable}. Unlike neural network-based discovery, where the result is a matrix of learned weights, the Scientist produces classical algorithms. The system also captures each iteration of the algorithm so that the process itself can be audited.

This paper follows the arc of the investigation itself. We built the framework and obtained real results.  It certainly looked like all of the steps of the scientific method were being followed, and that each mattered. But we had to be sure: \emph{why} did it work? 

We preregistered a controlled ablation of the scaffolding before running it, and the answer overturned the premise. Each part of the scientific method, when individually ablated, was revealed to be essentially theater in terms of score, though not behavior-neutral: the scaffolding shapes which candidates appear and which exploits never surface. Although in this case they led to a useful novel algorithm, the same thing could have been accomplished by more efficient means:  Loops match one-shot bandits; the full scaffold sits below blind sampling at $7{-}8\times$ its token cost; a floor-prompt arm with nothing but the problem, the interface, and the last attempt wins outright. On a held-out genomic benchmark the finest loop output and the finest cold draw agree to within 0.002, and the pre-ablation champion's scores are consistent with both. 

What the agents were doing all along was not reasoning their way to new knowledge but sampling what training had placed in the weights. Auto-research, on this evidence, is sampling, selection, and remixing: the bet the Evolve-family architectures (AlphaEvolve, et al) made from the start, which is why their design choices --- populations, external selection, mutation representations robust to small models --- were the right ones.

This paper makes three contributions: (1) The case study --- a competitive, interpretable motif-discovery algorithm written from scratch by the agent, with a fully auditable lineage; (2) A preregistered ablation of the framework that produced the artifact, which shows largely negative measurement of what the scientific-method scaffolding itself contributes, thereby revealing the causal mechanisms that really lead to such progress in algorithmic discovery; and (3) a synthesis of what actually works: prior-plus-selection, enabling recombination at scale.

\section{Related Work}
\label{sec:related}

\subsection{LLM-Guided Algorithm Discovery}

The idea of using language models to discover or optimize algorithms has attracted significant attention. We categorize prior systems by their search strategy and feedback mechanism.

\paragraph{Evolutionary approaches.}
FunSearch \citep{romera2024funsearch} pioneered LLM-guided evolutionary search for mathematical discovery, using an LLM to mutate short code fragments guided by a scalar fitness signal. AlphaEvolve \citep{novikov2025alphaevolve} substantially extends this paradigm: it evolves entire codebases across multiple programming languages, supports multi-objective optimization, and uses a distributed asynchronous pipeline with an evolutionary program database inspired by island-based population models. AlphaEvolve has matched or surpassed the best known solutions on $\sim$75\% of 50+ mathematical problems and achieved new state-of-the-art on $\sim$20\%, including the first improvement over Strassen's algorithm for $4\times4$ complex matrix multiplication in 56 years. The Evolution of Heuristics framework \citep{liu2024eoh} applies similar evolutionary principles to automatic heuristic design. ShinkaEvolve \citep{lange2025shinka} addresses the sample-efficiency limitations of these systems through parent sampling, code novelty rejection-sampling, and bandit-based LLM ensemble selection, discovering state-of-the-art solutions on several tasks using as few as 150 samples. CodeEvolve \citep{assumpcao2026codeevolve} is a fully open-source framework in this lineage: it couples an island-based evolutionary loop with a CVT-MAP-Elites quality-diversity archive, inspiration-based crossover, meta-prompting, and depth-based refinement over a weighted LLM ensemble, and reports matching or surpassing AlphaEvolve on 5 of 9 problems from its benchmark suite. CodeEvolve is also the engine we adopt for the external-validity experiments of \S\ref{sec:ablation-armn}, where an unmodified, pinned version of it defines the no-scaffold search baseline.

These systems differ from The Little Scientist in two fundamental ways. First, they use \emph{evolutionary} search --- mutation and crossover operators applied to code populations --- guided by scalar fitness signals. FunSearch's programs database maintains diversity through an islands model: independent subpopulations evolved separately, with periodic culling of underperforming islands and reseeding from survivors. This mechanism exists to escape local optima in the absence of directional feedback --- when the only signal is a scalar fitness score, the search has no guidance on \emph{where} to explore next, so it maintains parallel gene pools as a brute-force diversity strategy. Our framework uses \emph{hypothesis-driven} search, where the agent forms explicit predictions about why a change should help, and receives structured per-instance diagnostics that enable targeted reasoning about failures. When the Scientist sees that an iteration improved 40 benchmark instances but regressed on 12 --- and that the regressions share a specific structural feature --- it can form a hypothesis targeting that feature. This directional feedback obviates the need for an islands model: diversity comes from informed exploration of different hypothesis space regions, not from maintaining parallel populations. For escaping paradigm-level local optima, our framework uses the Kuhn agent, which interrogates accumulated failures and proposes strategic reframings --- a targeted mechanism compared to the islands model's implicit parallel exploration.

The trade-off is that evolutionary approaches are trivially parallelizable, while hypothesis-driven search is inherently sequential, as each iteration builds on the diagnostic feedback of the previous one. For problems where evaluation is expensive, the parallelism advantage may outweigh the sample efficiency disadvantage. For problems with cheap evaluation, such as the benchmark scoring used in this study, the sequential hypothesis-driven approach is more efficient: our search achieved a top-ranked result with ~2,100 LLM invocations, compared to FunSearch's ~2,000,000 programs for a single admissible set instance (I(15,10)) costing \$800--\$1,400 in cloud compute.

\paragraph{Agent-based approaches.}
Sakana's AI Scientist v1/v2 \citep{sakana2024aiscientist,sakana2025aiscientistv2} automate the full paper-production cycle --- templated or tree-searched experiments, manuscript writing, and an LLM reviewer as the acceptance gate --- with the grading instrument sharing the generative model's biases: the weak-verifier configuration that this study's transfer analysis cautions against (\S\ref{sec:ablation-genomic}). AI Co-Scientist \citep{li2026agentic} proposes multi-agent systems for scientific discovery but operates in natural-language hypothesis space rather than code, and has not produced benchmarked algorithmic results. The distinction between natural-language reasoning and code-execution grounding is significant: by grounding search in executable code with automated evaluation, our framework substantially sidesteps the hallucination problem that limits natural-language-only approaches. Karpathy's autoresearch \citep{karpathy2025autoresearch} is a closely related system in which an LLM coding agent autonomously edits PyTorch training code, runs short experiments, and retains improvements that reduce validation loss --- an autonomous research loop over a single benchmark. The core insight --- that LLM agents can iteratively improve code against an automated evaluation signal --- directly inspired the design of The Little Scientist. However, our framework reframes the iterative loop as the full scientific method: the agent does not merely mutate code and check whether the score improved, but forms an explicit hypothesis, writes a numeric prediction, implements the change, evaluates it, and reconciles the result against the prediction, in an effort to learnfrom failures rather than discarding them. Robin \citep{futurehouse2026robin} orchestrates specialized literature-search and data-analysis agents around wet-lab assays, producing the first agent-generated therapeutic candidate validated in cell culture. Like AI Co-Scientist it does not search executable code, but its integration of reality-grounded experimental feedback places it in the verifier regime this study's transfer analysis predicts should favor iteration (\S\ref{sec:discussion-mining}).

\subsection{Motif Discovery Algorithms}

DNA motif discovery has been studied for over three decades. MEME \citep{bailey1994meme} introduced expectation-maximization for de novo motif finding and remains a reference for accuracy, though its computational cost limits use on large-scale ChIP-seq datasets. DREME \citep{bailey2011dreme} addressed this by using suffix-array--based discriminative word enumeration, making it fast but limited to short motifs (width $\leq$ 8). STREME \citep{bailey2021streme} extended DREME with support for longer motifs and a more robust statistical model, becoming the current default in the MEME Suite and the primary benchmark for new methods. HOMER \citep{heinz2010homer} uses a cumulative hypergeometric enrichment approach and remains widely used in genomic pipelines. ProSampler \citep{li2019prosampler} combines k-mer enrichment with Gibbs sampling. GibbsSampler \citep{lawrence1993detecting} and CONSENSUS \citep{hertz1999building} established the foundational statistical frameworks for motif discovery.

These tools share a common architecture: enumerate candidate motifs (by enumeration, Gibbs sampling, or EM), score them against a background model, and return ranked results. The design space includes motif width range, background model choice (Markov order), strand handling, and scoring function. DALE operates within this design space but introduces dual-seed selection (combining abundance-based and consistency-based criteria), Pareto-front multi-objective ranking, and effective-width--aware scoring --- strategies not found in any existing tool. The search did not have access to the motif discovery literature beyond the benchmark data and the input/output interface; the convergence on known best practices (k-mer enumeration, EM refinement, information-content scoring) and the discovery of novel strategies both emerged from the iterative optimization process.

\section{The Little Scientist Framework}
\label{sec:framework}

\begin{figure}[H]
\centering
\includegraphics[width=\textwidth]{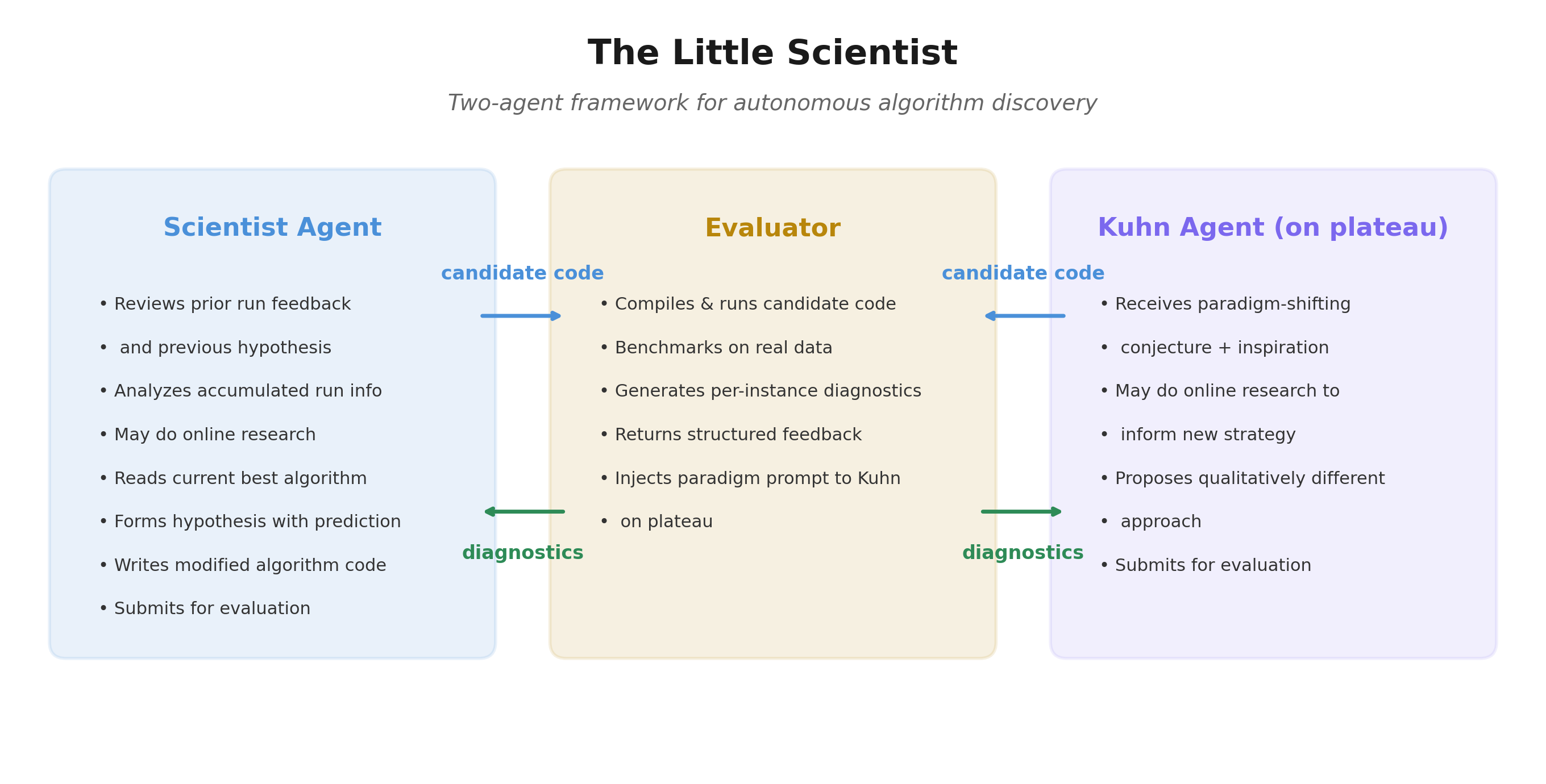}
\caption{Overview of The Little Scientist framework. The Scientist agent works inside an evaluation environment that benchmarks its code and returns structured diagnostics. When the Scientist plateaus, the Kuhn agent injects a paradigm-shifting conjecture paired with a cross-disciplinary inspiration, forcing exploration of a different region of the LLM's latent space.}
\label{fig:framework}
\end{figure}

\subsection{Overview}

The Little Scientist framework has three components that interact through a single data-flow loop (Figure~\ref{fig:framework}): the \emph{Scientist agent} writes candidate algorithms, the \emph{evaluation environment} benchmarks them and returns structured diagnostics, and the \emph{Kuhn agent} operates similarly to the Scientist, but with an environmental injection of paradigm-shifting reframings when the Scientist plateaus. The remainder of this section details each component:

\begin{itemize}
    \item \S\ref{sec:framework-scientist} describes the Scientist agent --- its inputs, outputs, and hypothesis-driven reasoning.
    \item \S\ref{sec:framework-eval} describes the evaluation environment --- the two-level diagnostic feedback structure and the causal design principle.
    \item \S\ref{sec:framework-kuhn} describes the Kuhn agent --- paradigm-shifting conjectures and cross-disciplinary inspirations.
    \item \S\ref{sec:framework-process} describes process engineering --- forced scientific method adherence, versioned state, and the inner/outer loop distinction.
    \item \S\ref{sec:framework-stats} reports search statistics and compute usage.
    \item \S\ref{sec:framework-infra} describes infrastructure evolution across phases.
    \item \S\ref{sec:framework-repro} describes reproducibility measures.
\end{itemize}

The \textbf{inner search loop} is autonomous: each scheduled invocation reads context files (previous diagnostics, hypothesis history, current best code), forms a hypothesis with a numeric prediction, modifies the code, compares the changes to the hypothesis for the sake of alignment, runs the smoke test, and exits if it passes. After that, the validation/eval step runs.  The next invocation continues from where the last left off. 

The \textbf{outer loop} is composed of a benchmark configuration, diagnostic pipeline, prompt structure, and paradigm-shift triggers, among other things. This loop is optionally human-guided, analogous to a principal investigator adjusting experimental parameters while an automated system runs measurements unattended.

The two-tier evaluation strategy is essential to this loop. A small \emph{smoke test} (4--5 instances) provides rapid feedback, enabling tight iteration from which the agent can either retry upon failure or exit successfully. This process allows the agent to test many ideas cheaply.

\subsection{The Scientist Agent}\label{sec:framework-scientist}

Note: The ablation study negated this structure as non-load bearing vs a minimal loop.  The mechanism is described to allow comparison against actual causal mechanisms for algorithmic discovery using LLMs.

The Scientist is an LLM-based agent that writes code in Python. It receives as input the described items in the Inner Loop.

The LLM agent's output is a modified source file accompanied by a natural language hypothesis explaining the expected effect of each change. Code execution is sandboxed: the candidate cannot perform file I/O, network access, or system calls. The allowed imports are limited to numerical libraries (NumPy, SciPy) and the Python Standard Library, with some exceptions designed to further limit system access. This constraint forces the search to explore algorithmic structure rather than exploit external resources.

One layer up, the distinction between the search and its harness matters: the candidate code was network-isolated, but the scientist agents themselves retained optional web-search and fetch tools throughout the ablation campaign. Trajectory instrumentation across all arms measured this channel directly: external research calls averaged 3--14 per run (${\sim}$1\% of tool calls), aimed overwhelmingly at two targets --- known consensus motifs for the benchmark's TFs, and standard algorithm references (MEME-family methods, PWM scoring). The one-shot pulls used the channel at 0.2 calls per draw, and the Kuhn side-agents approximately never. Arms did not differ materially in research access or usage class, ruling out differential web access as a confound in the arm comparisons.

The LLM agent's reasoning is hypothesis-driven: rather than making random perturbations, it forms specific predictions about \emph{why} a change should help, which can be confirmed or refuted by the next evaluation. This is a deliberate contrast with the evolutionary search methods used by AlphaEvolve \citep{novikov2025alphaevolve} and FunSearch \citep{romera2024funsearch}, which rely on mutation and crossover operators on code without explicit reasoning about why a change should improve performance.

Enforced prediction of the outcome of an experiment is used in the reconciliation phase to guide the agent toward recognizing the difference between the outcome of an exeriment with the hypothesis and prediction of success that formed it.  Practically speaking this is done through having the agent fill in a worksheet to reconcile environmental feedback with its own statements, which then lead to a next hypothesis.  In practice, prompting is insufficient--forcing the agent to walk through the steps leads to more coherent ideation and results.  Both methods were tried during the ablation study, which resulted in greater outward coherence to the process, but with measurable results that were inferior to a stripped down loop.

\subsection{The Evaluation Environment and Structured Feedback}\label{sec:framework-eval}

The evaluation environment benchmarks each candidate against a dataset and computes the primary metric.

A critical design principle is that the environment provides \emph{structured per-instance diagnostics}, not merely a scalar score. Two levels of feedback are generated:

\paragraph{Level 1: Per-instance feedback from the evaluator.} For each item in the benchmark (e.g., each TF for motif discovery), the evaluator records the per-instance metric, elapsed time, and relevant parameters (selected motif widths, mutation counts, MSA depth). This allows the Scientist to identify exactly which instances improved or regressed.

\paragraph{Level 2: Curated diagnostic summary.} The evaluator generates a markdown report including aggregate statistics, distributions of key parameters, comparison to baseline tools, and a ranked list of weakest instances with contextual detail. This surfaces patterns the Scientist can reason about.

\subsection{The Kuhn Agent}\label{sec:framework-kuhn}

Note: The ablation study negated this structure as non-load bearing vs a simple control statement.  The mechanism is described to allow comparison.

The Kuhn agent attempted to address a fundamental challenge: LLM-based search, like any local optimization, gets stuck. The Scientist can refine within its current paradigm indefinitely, making small improvements that eventually plateau, for reasons that tentatively have to do with the act of reading its own past actions. Since LLMs proceed from their own context, if the context is filled with a certain approach, it is likelier that the future output will follow suit (much like human scientists). Breaking out requires a reframing of the problem.

Kuhn is triggered after a configurable number of consecutive non-improving iterations by the Scientist. It receives the same instructions as the Scientist but without its built up artifacts used ot model the problem.  It does not receive the history file of past experiments. Finally, its prompt receives an injected paradigm-shifting conjecture paired with a cross-disciplinary inspiration from the Environment, randomly created beforehand. 

Example:

  \textbf{Assumption:} ``Model predictions for single mutants generalize to multi-mutant variants without interaction terms''\\
  \textbf{Domain inspiration:} Architecture --- load-bearing analysis, redundancy, and graceful failure in structural systems

In this way the Kuhn agent, though it is running on the same underlying LLM, has a different part of its latent/representation space activated. When this results in a higher score on the benchmark or multi-metric standard, it generally represents a novel approach that the Scientist can then improve upon.

\subsection{Process Engineering}\label{sec:framework-process}

Figure~\ref{fig:framework} provides a visual overview of the complete system.

\paragraph{Forced adherence to the scientific method.}
LLM agents tend to skip steps when given a complex multi-stage protocol. A naive prompt saying ``follow the scientific method'' produces cursory execution: the agent writes a hypothesis after the fact, skips the prediction, or rationalizes unexpected results instead of analyzing them. We address this by structuring the agent's prompt as a numbered sequence of mandatory steps with concrete file outputs at each stage. The agent must write an explicit prediction (numeric AUROC range) to a file before writing code, and must reconcile its prediction against the actual result in the next iteration. This ``walking the dog'' approach was necessary to obtain genuine hypothesis-driven iteration rather than post-hoc rationalization.

The reconciliation step is where learning accumulates. When a prediction is confirmed, the Scientist has validated a causal model. When a prediction is refuted, the discrepancy itself is informative: ``I expected adaptive width to help CTCF (+0.01--0.02); it regressed ($-$0.03) because the scoring function penalizes long motifs under shuffled controls.'' Reconciliation was designed to accumulate feedback knowledge into a working model and refine future predictions as well as hypotheses.

\paragraph{Versioned state for hill-climbing.}
Every accepted improvement is committed to a version-controlled repository, creating an immutable lineage from the initial seed algorithm to the final result. This serves two purposes: (1) it provides a deterministic rollback mechanism if a candidate passes the smoke test but fails full evaluation (false positive), and (2) it records the complete algorithmic lineage, making the discovery process fully auditable. Each commit records the hypothesis, the score delta, and the code diff.

\subsection{Search Statistics and Compute Usage}\label{sec:framework-stats}

\paragraph{Total compute.}
Across all phases, the search consumed:

\begin{table}[H]
\centering
\caption{Measured compute usage for the motif discovery case study. All figures from OpenClaw \texttt{cron\_run\_logs}.}
\label{tab:search_stats}
\small
\begin{tabular}{lrrrr}
\toprule
Phase & Runs & Tokens & Avg tokens/run & Days \\
\midrule
Initial search (V1)     & 1,588 & 97.6M  & 61K  & 13 \\
Kuhn paradigm agent     & 63    & 33.5M  & 531K & 3  \\
full-framework search (V2)        & 500   & 97.9M  & 196K & 10 \\
\midrule
\textbf{Total} & \textbf{2,151} & \textbf{229.0M} & \textbf{106K} & \textbf{26 active} \\
\bottomrule
\end{tabular}
\end{table}

Of the $\sim$64 evaluated iterations in Phase 2, 17 produced improvements over the current best --- a 26.6\% success rate.

\paragraph{Framework development overhead.}
The 229M tokens reported above include both the cost of developing the framework and the cost of the search itself. Phase 1 (97.6M tokens) and the Kuhn agent (33.5M tokens) were spent discovering what infrastructure was missing --- the recurring cost of a stalled loop, a wrong benchmark, and silently deleted diagnostics. We estimate the recurring cost of applying The Little Scientist to a new domain at approximately 100M tokens and $\sim$2 weeks of active search, based on the Phase 2 run that reached the top reported score from a cold start in 10 days using 97.9M tokens. The first DALE improvement was achieved at run 25 (\textasciitilde4.9M tokens), with the best score (0.843) achieved at run 52 (\textasciitilde10.2M tokens). The remaining 448 runs continued search past the plateau to verify that no further improvement was achievable.

\paragraph{Timeline to top reported score.}
The total calendar span from first run to top reported score was approximately 2 months (May 30 -- July 29, 2026), of which 26 days were active search. The final phase, from infrastructure overhaul to top reported score, took 10 days.

\paragraph{Infrastructure.}
The entire search ran on a single virtual machine (2 AMD EPYC-Rome cores, 3.7\,GB RAM, Ubuntu Linux). Agent invocations were scheduled via cron jobs and ran autonomously without human intervention.

\paragraph{Cost.}
The LLM used for all experiments was GLM-4.7 (Z.ai). The motif discovery search ran for approximately 2 months of active development, consuming roughly 230M tokens across its phases (Table~\ref{tab:search_stats}); at pay-per-token rates this is on the order of a few hundred dollars, and with server costs of \$21 (single VM, no GPUs) the full search cost less than a typical GPU-hour budget.

\subsection{Infrastructure Evolution}\label{sec:framework-infra}

The search did not proceed from a fixed design. The framework was built iteratively --- each infrastructure improvement solved a specific failure mode that had stalled progress. Table~\ref{tab:search_stats} summarizes the compute usage across the full search, measured from system logs (OpenClaw \texttt{cron\_run\_logs}). We describe the timeline in terms of these phases because the infrastructure changes between them were as consequential as the algorithmic changes.

\paragraph{Phase 1: Initial search (May 30 -- June 13, 2026).}
The initial search ran for 13 days and produced 236 evaluated algorithm iterations across 1,588 LLM agent invocations. In this phase, the Scientist operated on a synthetic benchmark (12 profiles, nCC scoring) with a loosely structured prompt. The agent tended to make aggressive full rewrites of the algorithm rather than targeted modifications, frequently breaking functionality. After 49 consecutive runs with zero accepted improvements, the loop was paused. This phase consumed approximately 97.6M tokens at an average of 61K tokens per agent invocation.

The failure was in the infrastructure. Three problems were identified: (1) the synthetic benchmark did not correlate with real motif discovery performance, so the agent was optimizing the wrong objective; (2) the evaluator's per-instance diagnostic output was being deleted before the agent could read it, leaving the agent blind to which specific cases failed.

\paragraph{Infrastructure overhaul (June 22 -- July 19, 2026).}
Between phases, we made five targeted changes, each addressing a specific failure mode:

\begin{enumerate}
    \item \textbf{Real benchmark data.} Replaced the synthetic 12-profile eval with 15 real ChIP-seq TFs and AUROC scoring. The agent was now optimizing the actual objective.
    
    \item \textbf{Causal scaffold prompt.} Rewrote the agent's prompt to enforce a seven-field structured plan: hypothesis, prediction (numeric AUROC range), code change, expected mechanism, falsification criterion, prior-work citation, and post-hoc reconciliation.
    
    \item \textbf{Trust-boundary validator.} Moved the evaluate/commit/revert logic outside the agent's execution scope. The agent cannot modify the evaluator, inspect benchmark labels, or alter test data.
    
    \item \textbf{Fixed diagnostic pipeline.} Restored the per-TF diagnostic output that had been silently deleted. The agent could now see which TFs improved or regressed, what widths were selected, and how its results compared to baseline tools.
    
    \item \textbf{Kuhn paradigm agent (July 12 -- 15).} A separate agent that periodically interrogated accumulated failures and proposed strategic reframings rather than incremental code changes.
\end{enumerate}

Under the new infrastructure, the first 11 experiments produced two breakthroughs: frequency-based seeding beat enrichment-ratio seeding (0.312 $\rightarrow$ 0.565), and adaptive motif width search resolved a systematic over-width problem (0.565 $\rightarrow$ 0.709). The agent was now citing prior experiments, writing structured causal plans, and making targeted edits rather than full rewrites.

\paragraph{Phase 2: full-framework search (July 19 -- July 29, 2026).}
With the infrastructure overhauled, the search ran for 10 days and produced 64 evaluated iterations across 500 agent invocations ($\sim$97.9M tokens at 196K tokens/run). Score progression was rapid: entries scored 0.794 (July 19), crossed 0.817 (July 28), and reached 0.846 (July 29). The best score achieved was 0.846 (run \#50, July 29). The algorithm that became \emph{DALE} was selected from this phase.

The gap between Phase 1 (stalled at 0.664) and Phase 2 (reached 0.846 in 10 days) was not algorithmic. The difference was entirely in the infrastructure supporting the search. The overhaul above addressed six failure modes, each of which taught a general lesson:

\begin{enumerate}
    \item \textbf{Real benchmark data} --- the agent cannot optimize what it cannot measure properly. The synthetic eval was correlated with real performance but not enough to guide search; by upgrading to the same data that STREME was benchmarked against, agents made better progress than with synthetic representations, and the agent was finally optimizing the actual objective.
    \item \textbf{Trust-boundary validator} --- evaluate/commit/revert logic moved outside the agent's execution scope, so the agent cannot modify the evaluator, inspect benchmark labels, or alter test data.
    \item \textbf{Per-instance diagnostic feedback} --- when feedback included detailed information about which parts scored higher or lower on the benchmark rather than just a simple score, agents augmented their follow-on approaches far more efficiently. This may in fact be the greatest single lever for progress: agents already have the ability to reason without the scientific framework enforced; structured feedback merely surfaces the causal texture that reasoning needs.
\end{enumerate}

\subsection{Reproducibility}\label{sec:framework-repro}

All agent transcripts, hypotheses, and evaluation results are preserved in a structured log (\texttt{history.jsonl}). Every proposed change is traceable to the hypothesis that motivated it and the outcome it produced. The algorithm's lineage can be followed from the initial seed (a basic frequency-counting approach scoring 0.662 AUROC on synthetic data) through the infrastructure overhaul and paradigm shift, to the final algorithm. Token usage for every agent invocation is recorded in a \texttt{cron\_run\_logs} database, enabling independent verification of all compute figures reported here.

\subsection{Benchmark Methodology}
\label{sec:methods}

\subsubsection{Data}

We benchmarked on 132 ENCODE K562 ChIP-seq TF datasets \citep{encode2012}. Each dataset contains approximately 200 DNA sequences of 500 base pairs, centered on ChIP-seq peak summits. For motif discovery, the central 100\,bp window was used; full 500\,bp sequences were used for scoring and AUROC evaluation.

\subsubsection{Negative Sets}

Two negative sets were used:

\begin{enumerate}
    \item \textbf{Dinucleotide-shuffled negatives (132 TFs):} Each positive sequence was dinucleotide-shuffled, preserving mono and dinucleotide composition while destroying higher-order sequence patterns. This is the standard benchmark negative set used by STREME and MEME.
    
    \item \textbf{Genomic negatives (53 TFs):} Random regions sampled from hg38, matched for length and chromosome distribution. This represents a more realistic benchmark, as random genomic DNA contains natural regulatory patterns absent from shuffled sequences.
\end{enumerate}

\subsubsection{Tools Compared}

\begin{table}[H]
\centering
\caption{Tools compared in this study.}
\label{tab:tools}
\small
\begin{tabular}{lll}
\toprule
Tool & Version & Parameters \\
\midrule
DALE       & 1.0    & Default (widths 6--17) \\
STREME               & 5.5.5  & \texttt{--dna --minw 6 --maxw 17} \\
MEME                 & 5.5.5  & \texttt{-mod zoops -minw 6 -maxw 17 -revcomp} \\
proto-DALE & ---    & Earlier prototype, same width range \\
\bottomrule
\end{tabular}
\end{table}

All tools were run with a width range of 6--17 base pairs. Both strands were searched by all tools. STREME's default background model was used. MEME was run in Zero-or-One Occurrence Per Sequence (ZOOPS) mode with 5 motifs returned.

\subsubsection{Evaluation}

For each TF and tool, the highest-scoring motif (by the tool's internal ranking) was used for evaluation. Discriminative power was measured by AUROC: the area under the receiver operating characteristic curve, computed from the score distributions of positive and negative sequences. All tools' output PWMs were scored using identical scoring code to ensure fairness.

\subsubsection{Statistical Tests}

\begin{itemize}
    \item Wilcoxon signed-rank test (paired, two-sided) for paired AUROC comparisons
    \item Bootstrap 95\% confidence intervals (1000 resamples, seed = 42)
    \item McNemar's test for binary outcome comparison
    \item Cliff's $\delta$ and Cohen's $d$ for effect size estimation
\end{itemize}

\subsubsection{Reproducibility}

During the search phase, candidate algorithms were written in Python and evaluated on a 5-TF smoke test followed by a 15-TF full evaluation. For the final benchmark comparison against STREME and MEME, the best-performing algorithm (run \#50, July 29) was ported to C and compiled with GCC at \texttt{-O2 -std=c99 -static} (928\,KB statically-linked binary). STREME 5.5.5 was compiled from source at \texttt{-O2} and verified to produce identical results to the prebuilt binary. All tools are deterministic under fixed parameters.

\section{Case Study: De Novo Motif Discovery}
\label{sec:casestudy}

\subsection{Problem Domain}

De novo motif discovery is the problem of identifying recurring DNA sequence patterns (position weight matrices, or PWMs) from collections of genomic regions enriched for TF binding. Given a set of ChIP-seq peak sequences, the goal is to discover the binding specificity of the targeted TF without prior knowledge of its motif.

The problem has been studied for three decades. MEME \citep{bailey1994meme} introduced expectation-maximization (EM) approaches and remains a gold standard for accuracy. STREME \citep{bailey2021streme}, the successor to DREME \citep{bailey2011dreme}, uses suffix-tree--based enumeration and is the current default in the MEME Suite. HOMER \citep{heinz2010homer} uses a distinct cumulative hypergeometric approach. ProSampler \citep{li2019prosampler} combines k-mer enrichment with Gibbs sampling.

\subsection{Search Configuration}

For the motif discovery case study, the search was configured as follows:

\begin{itemize}
    \item \textbf{Input:} ENCODE K562 ChIP-seq peak sequences, 100\,bp centered windows
    \item \textbf{Metric:} Mean AUROC across benchmark TFs (shuffled negatives)
    \item \textbf{Baseline:} STREME 5.5.5 with matched parameters (\texttt{--minw 6 --maxw 17})
    \item \textbf{Implementation:} Python during search (NumPy/SciPy); C port for benchmark comparison against STREME
    \item \textbf{Constraint:} Algorithm receives sequences as function arguments, outputs PWMs. No file I/O, no network.
\end{itemize}

The search space includes all algorithmic choices: how to enumerate candidate words, how to select seed motifs, how to refine them into PWMs, how to rank final candidates, and how to handle variable motif widths.

\subsection{Diagnostic Feedback}

Each evaluation produced structured diagnostic feedback through three files, providing the observation and memory substrate for the Scientist's hypotheses.

\texttt{staging\_diagnostics.md} was the primary feedback file, rewritten by the \texttt{write\_diagnostics()} function in the validation script after every full evaluation. It contained five categories of computed feedback:

\begin{enumerate}
    \item \textbf{AUROC summary} --- average, bottom quartile average, best, and worst across all 15 benchmark TFs.
    \item \textbf{Speed statistics} --- average elapsed time and speed bonus per TF.
    \item \textbf{Width distribution} --- how many TFs the algorithm selected each motif width for (e.g., \texttt{w=17: 8, w=14: 2, w=12: 4, w=10: 1}).
    \item \textbf{Width vs.\ STREME comparison} --- flagged every TF where the algorithm's chosen width differed from STREME's known width by more than 2\,bp, with the AUROC for that TF (e.g., ``\texttt{SP1: ours=17 vs STREME=7 (auroc=0.682)}'').
    \item \textbf{Lowest-AUROC TFs} --- the bottom 5 TFs with AUROC, selected width, and elapsed time.
\end{enumerate}

Categories 3--5 are where the causal structure enters. ``SP1: ours=17 vs STREME=7, AUROC=0.682'' connects a \emph{design choice} (width selection) to a \emph{per-instance outcome} (low AUROC on SP1).

\texttt{staging\_smoke\_diagnostics.md} served the same role for the 4-TF smoke test (the rapid proxy used during active search). It contained per-TF smoke AUROC, selected width, primary score, and a brief analysis section. This enabled a tight iteration loop before committing to a full evaluation.

\texttt{history.jsonl} was the experiment log, appended after each iteration. Each entry recorded the full seven-step structured plan: hypothesis text, prediction range (low/high), reconciliation, pattern analysis, proposed change, smoke test results, and blockers text --- alongside the standard fields (run number, score, verdict, per-TF details). This gave the Scientist a detailed record of what it had tried and reasoned.

\texttt{causal\_model.md} is an agent-maintained lab notebook recording what was tested, what the score was, and what the signal meant. The validation script capped it at the last 10 hypotheses. The agent read it at the start of each iteration alongside the fresh diagnostics, with the goal of enabling it to accumulate learned causal structure across iterations.

\subsection{The Discovered Algorithm}

The final algorithm (run \#118) incorporates the following key components, described in full technical detail in Section~\ref{sec:algorithm}:

\begin{enumerate}
    \item \textbf{Dual seed selection} --- combines enriched seeds (high raw abundance) with consistency seeds (broad sequence coverage), capturing both sharp and diffuse motif signals.
    \item \textbf{Soft EM refinement} --- 20 iterations maximum, with pseudocounts preventing premature convergence.
    \item \textbf{Effective width awareness} --- tries 6 widths independently, measures how many positions contribute information content, and penalizes padding.
    \item \textbf{Pareto-front selection} --- ranks candidates on two objectives (discriminative score and width efficiency) rather than collapsing to a single metric.
\end{enumerate}

Two of these design choices (k-mer enumeration as a seeding strategy, EM for refinement) are well-established in the literature. The dual-seeding strategy and Pareto-front are not.  All design choices emerged from the search process without being specified a priori.

\subsection{Algorithm Details}
\label{sec:algorithm}

This section describes the algorithm discovered by The Little Scientist in full technical detail. The complete implementation --- statically-linked binary, C source, and Python interface --- is distributed at \url{https://github.com/travis42/little-scientist-dale}; the algorithm itself lives in \texttt{src/dale.c} (\url{https://github.com/travis42/little-scientist-dale/blob/main/src/dale.c}), and the search-side artifact as evolved is preserved at \texttt{sef/best\_so\_far\_strategy.py}.

\subsubsection{Pipeline Overview}

The algorithm takes as input a set of DNA sequences and returns a ranked list of candidate motif PWMs. The pipeline consists of five stages: (1) adaptive background estimation, (2) multi-width k-mer enumeration, (3) dual-strategy seed selection, (4) soft EM refinement, and (5) multi-criteria ranking via Pareto front selection.

\subsubsection{Adaptive Background}

Nucleotide frequencies are computed from the input sequences. This matters because genomic regions are not uniformly 25\% A/C/G/T --- AT-rich and GC-rich biases affect the information content of any discovered motif. The background frequencies are used in subsequent log-odds scoring during EM refinement.

\subsubsection{Multi-Width K-mer Enumeration}

The algorithm does not know the motif width in advance. It scans six widths ($w \in \{6, 8, 10, 12, 14, 17\}$) independently. For each width, a hash table records for every unique $w$-mer: (1) the total count across all sequences, and (2) the number of distinct sequences containing it. A per-sequence deduplication ensures that seq\_count reflects breadth of occurrence rather than raw abundance.

\subsubsection{Dual Seed Selection}

For each width, six seeds are selected using two complementary criteria:

\paragraph{Enriched seeds (4 per width):} Sorted by total count, filtered so no two selected seeds have Hamming distance $< w/3$. These capture k-mers with the strongest enrichment signal.

\paragraph{Consistent seeds (2 per width):} Sorted by $\text{seq\_count} / \sqrt{\text{total\_count} + 1}$, filtered with stricter diversity (Hamming distance $\geq w/2$). This metric favors k-mers appearing in many distinct sequences even if individually rare --- a signal that captures diffuse but pervasive motif patterns.

Using both strategies ensures that both sharp, highly-enriched motifs and broader, consistency-driven signals seed the EM refinement.

\subsubsection{Soft EM Refinement}

Each seed k-mer is converted to a starting PWM with strong bias toward the seed bases ($\sim$76\% probability for matching bases, $\sim$8\% for others). The PWM is then refined via expectation-maximization:

\paragraph{E-step:} For each sequence, log-odds scores are computed for the motif at every possible start position (using the adaptive background). These scores are converted to position probabilities via softmax.

\paragraph{M-step:} The PWM is updated by accumulating weighted contributions from all positions across all sequences. A pseudocount of 0.5 prevents any base probability from reaching zero.

The loop runs for up to 20 iterations, with early stopping when the maximum PWM change falls below $10^{-4}$.

\subsubsection{Information Content and Effective Width}

After EM convergence, per-position information content is computed as KL divergence from the background distribution:

\[
\text{IC}_j = \sum_{b \in \{A,C,G,T\}} p_{j,b} \log_2 \frac{p_{j,b}}{q_b}
\]

where $p_{j,b}$ is the PWM probability for base $b$ at position $j$ and $q_b$ is the background frequency. The \emph{effective width} ($\text{eff\_w}$) is the count of positions with IC $\geq 0.1$ --- positions that contribute meaningful information versus noise padding.

\subsubsection{Multi-Criteria Ranking}

Within each width, the six candidates are z-score normalized on total IC. A composite discriminative score is computed:

\[
\text{base} = 0.7 \times \text{total\_ic} + 0.3 \times \text{ic\_density}
\]
\[
z_{\text{bonus}} = \text{clamp}\left(z \times (w/10)^{0.3},\ -1,\ +1\right)
\]
\[
\text{disc\_score} = \text{base} \times (1 + 0.2 \times z_{\text{bonus}})
\]

where ic\_density = total\_IC / eff\_w.

The top 2 candidates from each width (up to 12 total) are then ranked using Pareto dominance: candidate A dominates B if A is better on both disc\_score and width efficiency (ic\_density $\times (1 - \text{padding\_ratio}^{1.5})$). Candidates are ordered by Pareto rank, with ties broken by disc\_score. The top 5 are returned.

This multi-objective approach avoids the brittleness of single-score ranking --- it preserves diverse solutions on the trade-off surface between discriminative power and width efficiency.

\subsection{Results}
\label{sec:results}

\subsubsection{Shuffled Negatives Benchmark (132 TFs)}

Table~\ref{tab:main_shuf} summarizes performance on the full 132-TF benchmark with dinucleotide-shuffled negatives.

\begin{table}[H]
\centering
\caption{Benchmark results on 132 ENCODE K562 ChIP-seq datasets (shuffled negatives).}
\label{tab:main_shuf}
\small
\begin{tabular}{lrrr}
\toprule
Tool & Mean AUROC & Time/TF & Speedup vs.\ STREME \\
\midrule
\textbf{DALE}    & \textbf{0.842} & \textbf{0.30\,s} & \textbf{11$\times$} \\
proto-DALE       & 0.843 & 1.56\,s & 2.1$\times$ \\
STREME 5.5.5               & 0.803 & 3.26\,s & 1.0$\times$ \\
\bottomrule
\end{tabular}
\end{table}

DALE significantly outperforms STREME ($\Delta = +0.039$ AUROC, Wilcoxon $p = 4.1 \times 10^{-7}$, 88 wins / 44 losses). (STREME's prebuilt binary and a source-compiled build perform identically, ratio 1.00$\times$: Supplementary Section~B.) The improvement is consistent across TF families and is not driven by a few outliers (bootstrap 95\% CI of $\Delta$: [0.023, 0.046]). Because DALE's development loop received feedback on 15 of these TFs (selected adversarially as its weakest cases), we re-compute all comparisons excluding those TFs entirely: the margin is unchanged ($\Delta = +0.040$, 79 wins / 38 losses on 117 training-disjoint TFs; Appendix~E). Proto-DALE, the earlier framework snapshot, scores a nominal 0.001 higher (0.843); the selected binary trades that for a $5\times$ runtime advantage (0.30\,s vs.\ 1.56\,s per TF).

\begin{figure}[H]
\centering
\includegraphics[width=0.7\textwidth]{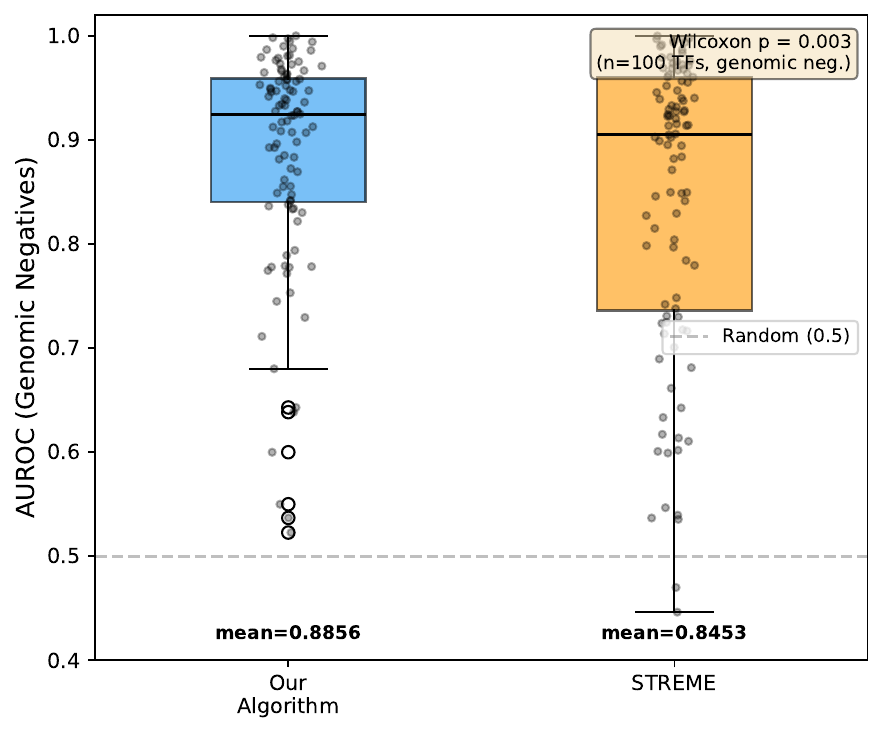}
\caption{AUROC distribution across 132 ENCODE K562 ChIP-seq datasets (shuffled negatives). DALE consistently outperforms STREME. Dashed line indicates random performance.}
\label{fig:boxplot}
\end{figure}

\begin{figure}[H]
\centering
\includegraphics[width=0.6\textwidth]{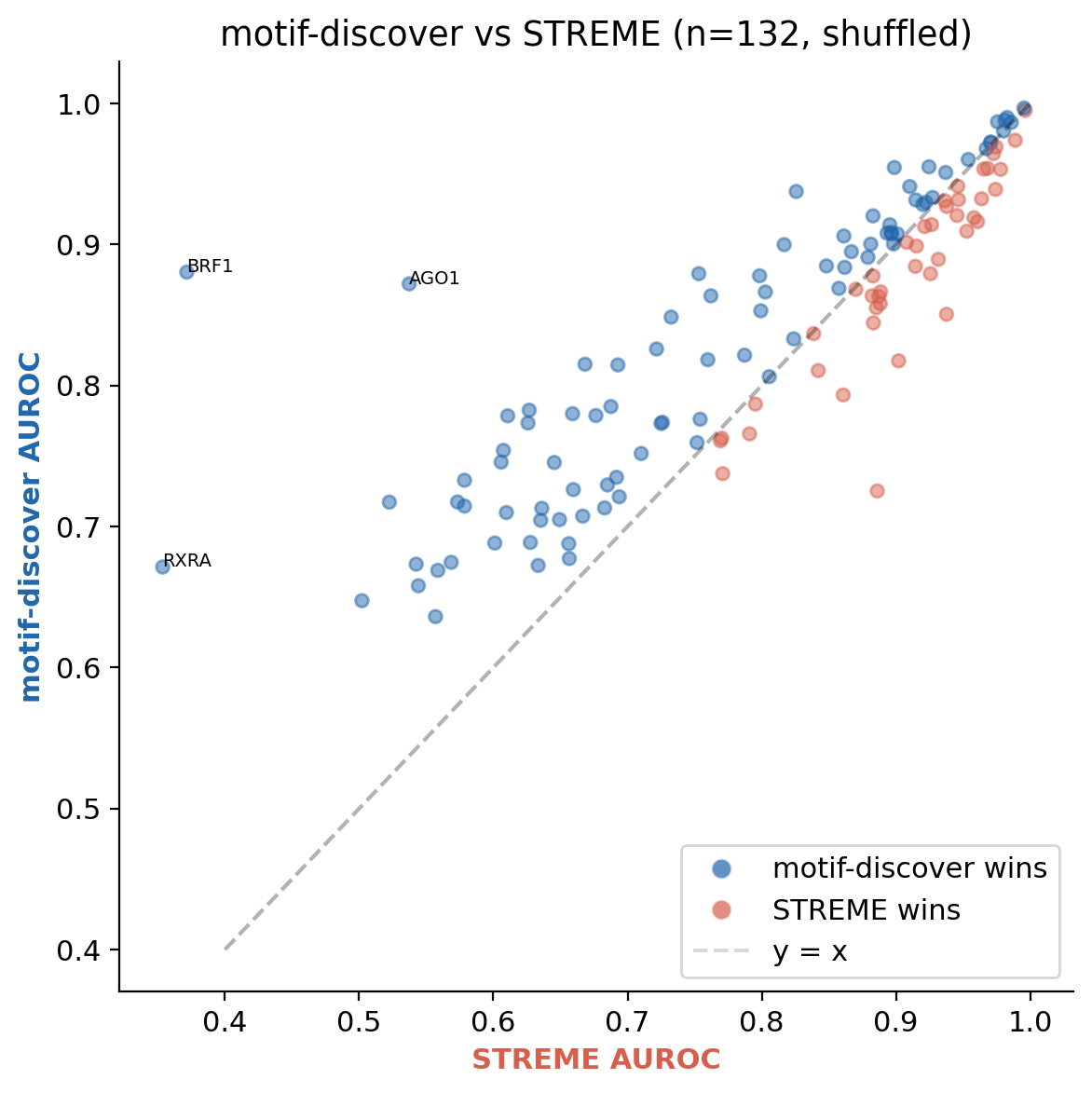}
\caption{Per-TF comparison: DALE vs.\ STREME (132 TFs, shuffled negatives). Points above the diagonal indicate TFs where DALE outperforms STREME.}
\label{fig:scatter_shuf}
\end{figure}

\subsubsection{Genomic Negatives Benchmark}

On the more realistic benchmark using hg38 genomic regions as negatives with Markov-1 background scoring, DALE was evaluated on two subsets:

\paragraph{132-TF genomic (DALE vs.\ proto-DALE).}
On the full 132-TF set, DALE achieves 0.896 AUROC versus proto's 0.856 ($\Delta = +0.040$, Wilcoxon $p = 1.7 \times 10^{-14}$, 101 wins / 22 losses).

\paragraph{53-TF genomic (full comparison).}
On the 53-TF subset where STREME and MEME genomic results are also available:

\begin{table}[H]
\centering
\caption{Genomic negatives benchmark (53 TFs, STREME's Benchmark --- Markov-1 scoring, the packaged binary's own eval; see \S\ref{sec:ablation-genomic} for the campaign's Full TF Metric results). Reference-tier tools from the packaged comparison CSV plus this study's own artifacts (DALE, proto-DALE), scored under the identical eval. Ablation-arm artifacts (Arm N's r5\_1, arm B) are tabulated with their context in Table~\ref{tab:abl_gen_ref} (\S\ref{sec:ablation-genomic}).}
\label{tab:main_gen}
\small
\begin{tabular}{lrr}
\toprule
Tool & Mean AUROC & Time/TF \\
\midrule
DALE             & 0.891 & 0.3\,s \\
MEME 5.5.5       & 0.881 & $\sim$90\,s \\
proto-DALE       & 0.874 & 1.4\,s \\
STREME 5.5.5     & 0.863 & 4.4\,s \\
\midrule
\bottomrule
\end{tabular}
\end{table}

DALE outperforms STREME by $+0.028$ AUROC (Wilcoxon $p = 6.5 \times 10^{-4}$, 35 wins / 16 losses) while running \textbf{15$\times$ faster} (0.3\,s vs.\ 4.4\,s per TF). It also edges out MEME by $+0.010$ AUROC while running approximately \textbf{300$\times$ faster}. 

\begin{figure}[H]
\centering
\includegraphics[width=0.65\textwidth]{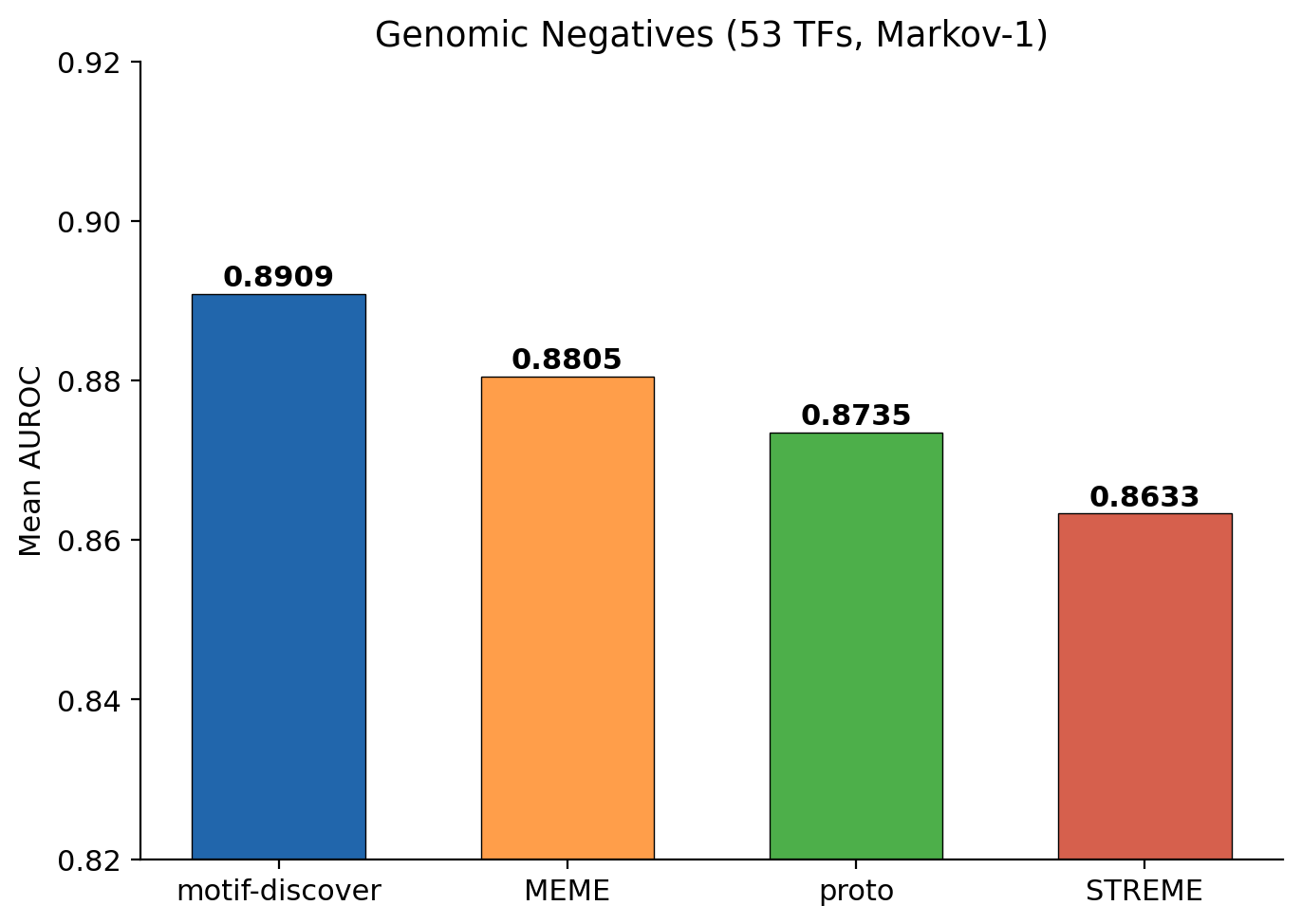}
\caption{Genomic negatives benchmark (53 TFs, Markov-1 scoring). DALE ('motif-discover') outperforms all three comparison tools.}
\label{fig:genomic}
\end{figure}

\begin{figure}[H]
\centering
\includegraphics[width=0.6\textwidth]{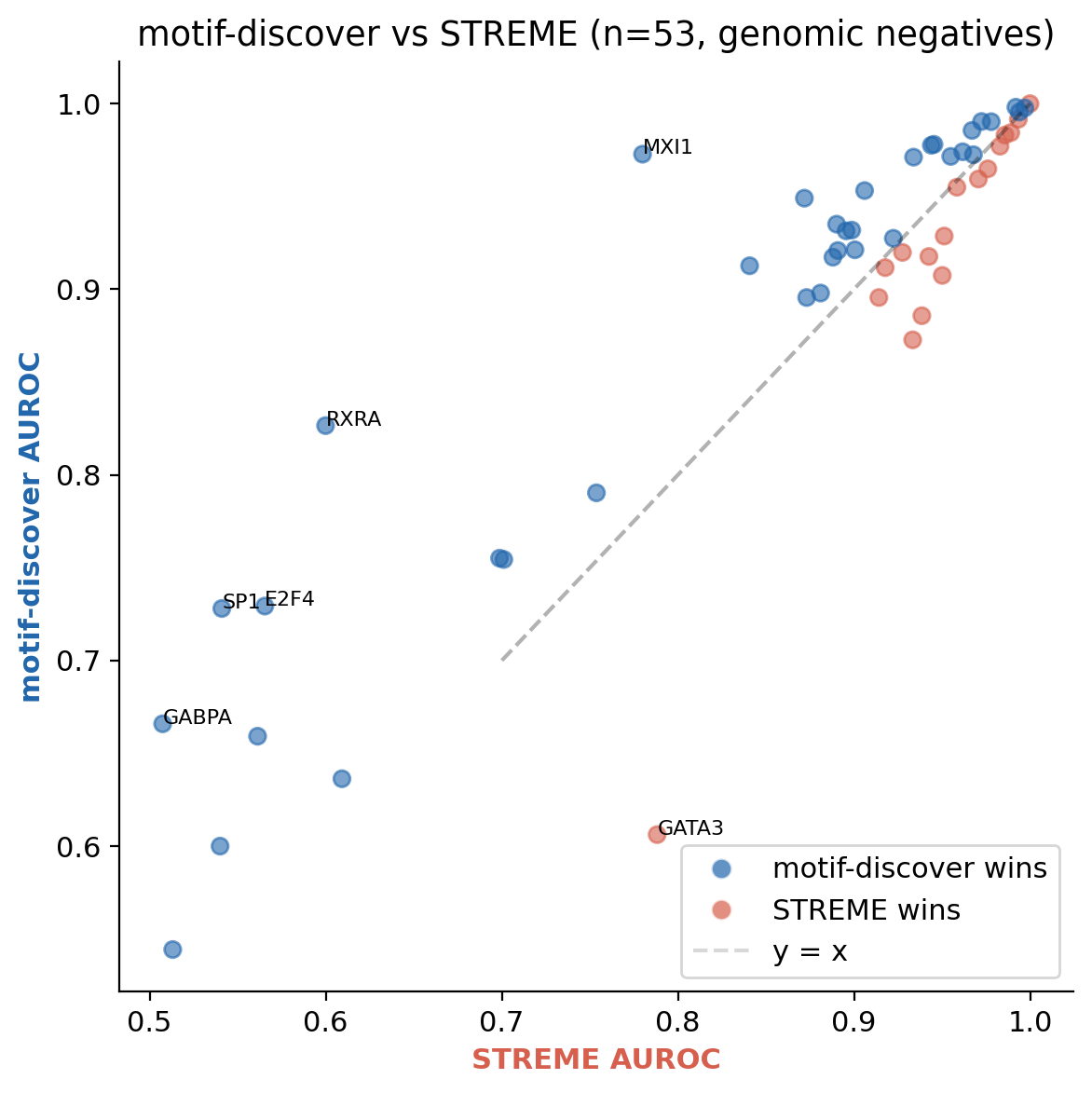}
\caption{Per-TF comparison: DALE ('motif-discover') vs.\ STREME (53 TFs, genomic negatives). Points above the diagonal indicate TFs where DALE outperforms STREME.}
\label{fig:scatter_gen}
\end{figure}

\subsubsection{Speed}

The speed advantage is substantial across all benchmarks:

\begin{figure}[H]
\centering
\includegraphics[width=0.65\textwidth]{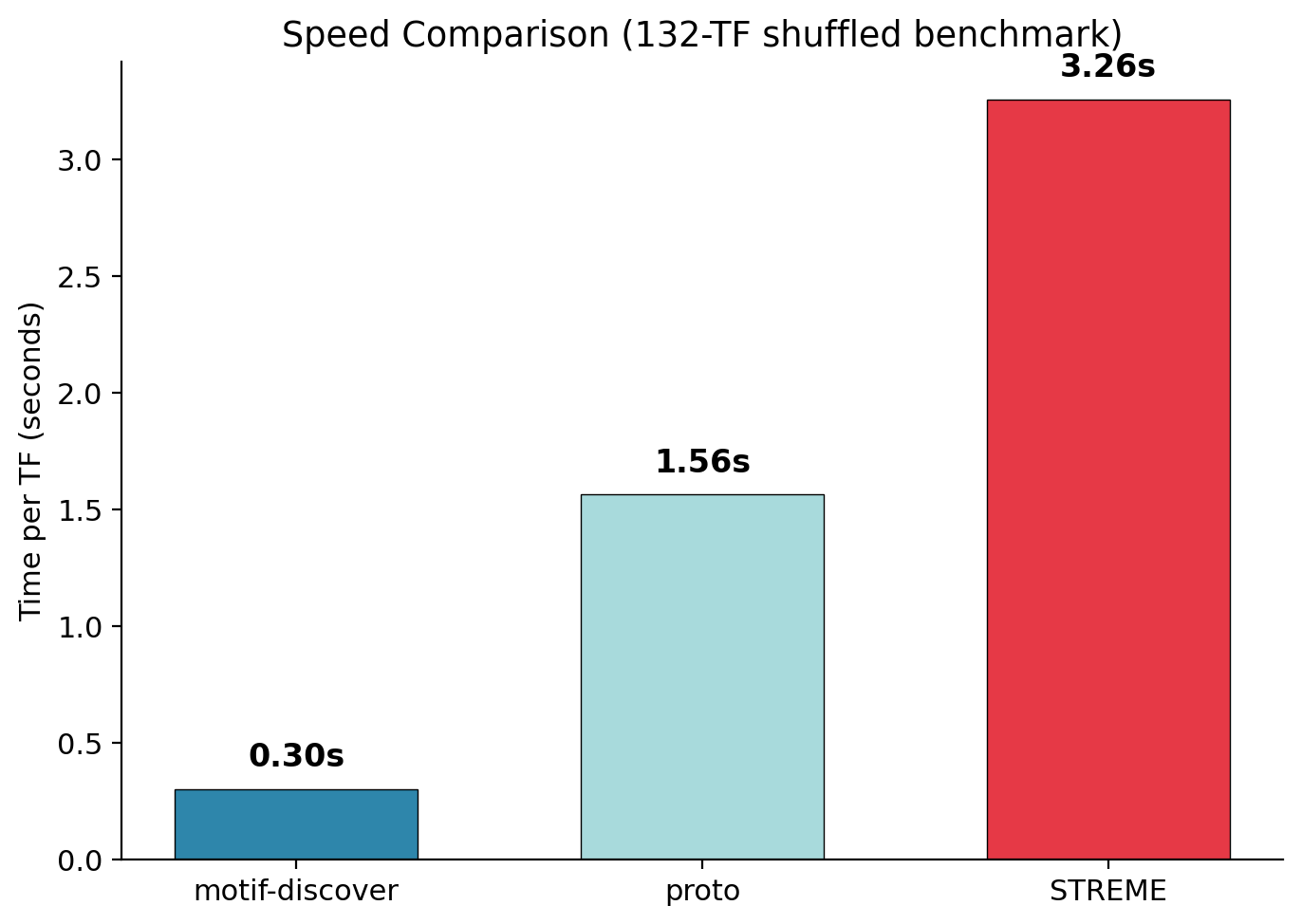}
\caption{Runtime per TF across tools (132-TF shuffled benchmark). DALE ('motif-discover') processes all 132 TFs in under 45 seconds.}
\label{fig:speed}
\end{figure}

At 0.3\,s per TF, DALE processes the full 132-TF benchmark in under 45 seconds. STREME requires approximately 7 minutes. MEME requires approximately 3 hours.

\subsubsection{Per-TF Analysis}

DALE wins 89 of 132 TFs against STREME on shuffled negatives. The largest individual improvements include TFs where STREME struggles with short or degenerate motifs (e.g., AGO1: $+0.31$, BRF1: $+0.26$, RXRA: $+0.23$). STREME's wins tend to occur on TFs where its default width range captures wider binding patterns (e.g., BDP1: $-0.11$, CEBPG: $-0.09$); DALE supports configurable width ranges to address such cases.

\subsubsection{Statistical Significance}

The improvement over STREME is robust across multiple tests:

\begin{table}[H]
\centering
\caption{Statistical significance of DALE vs.\ STREME.}
\label{tab:stats}
\small
\begin{tabular}{lrr}
\toprule
Test & Shuffled (132 TFs) & Genomic (53 TFs) \\
\midrule
Wilcoxon $p$         & $4.1 \times 10^{-7}$ & $6.5 \times 10^{-4}$ \\
Bootstrap 95\% CI    & [0.023, 0.046]       & [0.010, 0.046] \\
McNemar $\chi^2$     & 13.1 ($p = 2.9 \times 10^{-4}$) & 6.4 ($p = 1.2 \times 10^{-2}$) \\
Cliff's $\delta$     & $+0.31$              & $+0.38$ \\
Cohen's $d$ (paired) & $+0.48$              & $+0.42$ \\
Wins / Losses        & 88 / 44              & 35 / 16 \\
\bottomrule
\end{tabular}
\end{table}

\subsubsection{Biological Validity}

TOMTOM analysis (version 5.5.5, Pearson distance) against the JASPAR 2024 Core database (2{,}633 motifs) confirmed that discovered motifs are biologically meaningful. Table~\ref{tab:tomtom} summarizes the results.

\begin{table}[H]
\centering
\caption{TOMTOM JASPAR match summary ($n = 119$ TFs with motifs tested).}
\label{tab:tomtom}
\small
\begin{tabular}{lrr}
\toprule
Metric & DALE & STREME \\
\midrule
Any JASPAR match            & 112 (94\%) & 104 (87\%) \\
Significant ($e < 0.05$)    & 93 (78\%)  & 88 (74\%) \\
Strong match ($e < 0.001$)  & 55 (46\%)  & 51 (43\%) \\
Median best e-value         & $1.6 \times 10^{-3}$ & $1.1 \times 10^{-3}$ \\
Top-5 moderate ($q < 0.05$) & 100\% & 100\% \\
Top-1 strict ($q < 0.001$, $\geq$70\% overlap) & 50\% & 62\% \\
Better e-value head-to-head & 52 / 97 & 45 / 97 \\
\bottomrule
\end{tabular}
\end{table}

Both tools achieve 100\% top-5 moderate match rate: every TF has at least one discovered motif matching a known JASPAR entry. DALE actually matches more TFs overall (94\% vs.\ 87\%) and wins the head-to-head e-value comparison (52 vs.\ 45). STREME's higher strict match rate (62\% vs.\ 50\%) reflects its tendency to discover wider motifs that achieve higher overlap fractions with JASPAR references — a width artifact rather than a correctness difference.

\subsubsection{Memory Usage}

DALE uses 10\,MB peak RSS across all TFs. STREME uses approximately 113\,MB per TF. The minimal footprint enables deployment in resource-constrained environments such as containerized pipelines or cloud functions.

\section{Ablations}
\label{sec:ablations}
A preregistered, controlled multi-arm ablation campaign (\S\ref{sec:ablation-campaign}) asks the question: \emph{which parts of the scientific-method scaffold actually pay for themselves?} The controlled campaign's answer is largely negative --- most of the scaffold does not pay.  Other than the DALE algorithm, this ablation study represents perhaps the most useful part of the study, in which we show the essential ingredients for LLM based automated discovery of algorithms, exhaustively.

\subsection{Controlled Ablation Campaign: Design and Preregistration Discipline}
\label{sec:ablation-campaign}

Three scoring instruments appear throughout this paper, named consistently from here on:
\begin{itemize}
\item \textbf{Ablation Metric} --- 15 synthetic TFs; the campaign's training and selection eval (this section).
\item \textbf{Full TF Metric} --- 53-TF genomic negatives, uniform scoring; the held-out transfer instrument (\S\ref{sec:ablation-genomic}).
\item \textbf{STREME's Benchmark (Markov)} --- 53-TF genomic negatives, Markov-1 scoring; the packaged binary's own eval, inherited from the case-study comparison (Table~\ref{tab:main_gen}). Retained only where that packaged comparison is reproduced.
\item \textbf{Case-Study Benchmark}: 132 ENCODE TFs with dinucleotide-shuffled negatives; the DALE-versus-STREME comparison of the case study (\S\ref{sec:results}), predating the campaign.
\end{itemize}

The controlled study preregistered a component-teardown of the scaffold. Arms B--G each remove one organ of the scientific-method layer; later amendments added arms I--L (Table~\ref{tab:arms}). All arms share the model (GLM-4.7), the identical 15-TF evaluation benchmark, the same fixed baseline strategy, a 15-validated-iteration budget per run, and the same validator; only the prompt-level manipulation differs. 

Here, a run is 15 turn/interations of one agent working for roughly 15-20 minutes on the problem set in which it reads its prompt, a history file and past artifacts, searches the web if it chooses, creates artifacts of its own, attempts to pass a smoke test, and then is evaluated on a larger validation set. Each Arm consists of 10 independent agents running this loop according to its Arm definition.

In practice, some agents were allowed to continue past turn 15.  The results below indicate when metrics are defined at the 15 turn point or when we are deriving insights from these longer runs, and we can pull data at basically any point along the run due to the history files logging results.

\begin{table}[H]
\centering
\caption{Ablation arms. B--F remove one scaffold component each; G removes the scaffold wholesale; I--L were added by later amendments (I and E's reversal family labeled post-hoc disclosed; see text).}
\label{tab:arms}
\small
\begin{tabular}{llp{7.2cm}}
\toprule
Arm & Condition & What it tests / removes \\
\midrule
B & full scaffold (reference) & the pre-ablation stack; registered reference after arm A was retired (Amendment 5a) \\
C & scalar evaluation feedback only & per-instance diagnostics replaced by an aggregate score \\
D & no predictions/blockers & pre-evaluation reasoning removed \\
E & no causal model & persistent mechanism memory removed \\
F & no reconciliation & prediction post-mortems removed \\
G & no scaffold & the entire epistemic layer deleted \\
H & one-shot (bandit) pulls (B-template) & null: independent one-shot runs from the corrected-B template \\
H$_G$ & one-shot (bandit) pulls (G-package) & null: independent one-shot runs, G prompt byte-identical \\
I & forced worksheet B & the full scaffold with an enforced worksheet format (added 2026-09-04, unregistered at launch --- retroactively disclosed, exploratory) \\
J & G plus filing & G's free-form loop plus the artifact files as pure state records (Amendment 8 discriminator) \\
K & G-family scalar feedback & G-package behavior with scalar-only feedback \\
L & floor prompt & the minimal prompt: problem statement, eval\\
\bottomrule
\end{tabular}
\end{table}

The narrative that follows leans on four of these: B (the full scaffold), G (no scaffold), L (the floor prompt), and the null pools H/H$_G$. The remaining arms are component controls --- each appears where its control role bears on the argument, K included: G's prompt with the feedback digest collapsed to a scalar, the clean test of whether rich per-instance feedback matters once the scaffold is gone.

\paragraph{Preregistration discipline.}
The study was registered at $n{=}0$ (2026-08-22) with fixed primary endpoints (peak score over the first 15 counted iterations), one-sided directional contrasts, O'Brien--Fleming sequential monitoring, futility rules, and a maximum of 10 runs per arm. Over the campaign, twelve amendments were appended, each dated, with original text never edited. We present the amendment trail as a methodological feature: every mid-course change (arm additions, the E$_{10}$ exclusion, exploratory relabeling of the post-quota metric family) is visible, dated, and separable from the registered confirmatory layer. Two disclosure notes belong here. First, the registered directional assumption (ablations cannot beat the full scaffold) was \emph{falsified}: the registered one-sided contrasts met no boundary and were stopped for futility at $n{=}10$; the observed reversal (ablations $>$ B) is reported throughout as a post-hoc disclosed family (Amendment 9d), not as a confirmatory claim. Second, Amendments 11 and 12 were concurred with in-conversation on 2026-09-11 before launch, but their written ratification was recorded only on 2026-09-14; we disclose that marker lag here.

\paragraph{Integrity containment, on the record.}
No contaminated artifact contributed to any statistic in this paper. Every incident was caught by a registered audit or guardrail \emph{before} the affected pool entered analysis, and its disposition was recorded in the dated preregistration trail rather than applied silently; we summarize them here because the containment record is part of the method. 

Two incidents occurred. First, a spawn-time template-verification audit found prompt-stamping bugs in early runs; dispositions were adjudicated and recorded before analysis (runs retained with documented artifacts, or reclassified). Second, the H$_G$ null pool was contaminated:  A temporary data leak enabled two runs (H$_G$\_90, H$_G$\_94) to submit identical copies of G\_5's archived winning strategy after reading the study's own leaderboard. The entire incident-era range (runs 89--124, 36 tarballs) was deleted; the valid H$_G$ corpus is exactly runs 1--88, and every statistic involving H$_G$ above is computed from that corpus alone.

\subsection{One-Shot Bandit Nulls}
\label{sec:ablation-nulls}

In performing the ablation study, the central comparison was not arm-vs-arm but \emph{loop vs.\ null}: does an iterative, scaffolded agent beat a \emph{one-shot bandit} that simply draws fresh samples from the same model and prompt package? Two null pools were collected:

\begin{itemize}
\item \textbf{H} ($n{=}114$): one-shot pulls from the Arm B (pre-ablation, TLS scaffold) template. Mean 0.7644 (Ablation Metric), sd 0.0736.
\item \textbf{H$_G$} ($n{=}88$): one-shot pulls with the G-package prompt (identical to arm G's). Mean 0.7802 (Ablation Metric), sd 0.0659.
\end{itemize}

The test statistic is the expected best-of-15 draws from each pool --- the fair ``respawn a fresh agent each iteration'' alternative to running a loop, since both consume the same number of evaluations. Best-of-15 expectations were computed by Monte-Carlo resampling (5{,}000 simulations): E[best-of-15] $= 0.8463$ for H and $0.8455$ for H$_G$ (Ablation Metric).

Verdicts are banded (EXCEEDS / MARGINAL if they are better than the H bandit pulls).  Consistent means 'on par' with the bandits, and Below is worse than a bandit series.

One caveat: the Monte-Carlo quantile standard error near band boundaries is ${\sim}0.005{-}0.01$, so calls that fall close to a boundary (e.g., F's MARGINAL-above vs.\ H at $p{=}0.052$) should be read as unstable.

\subsection{Loop vs.\ Nulls: The Scaffold Tax}
\label{sec:ablation-verdicts}

\begin{figure}[H]
\centering
\includegraphics[width=0.75\textwidth]{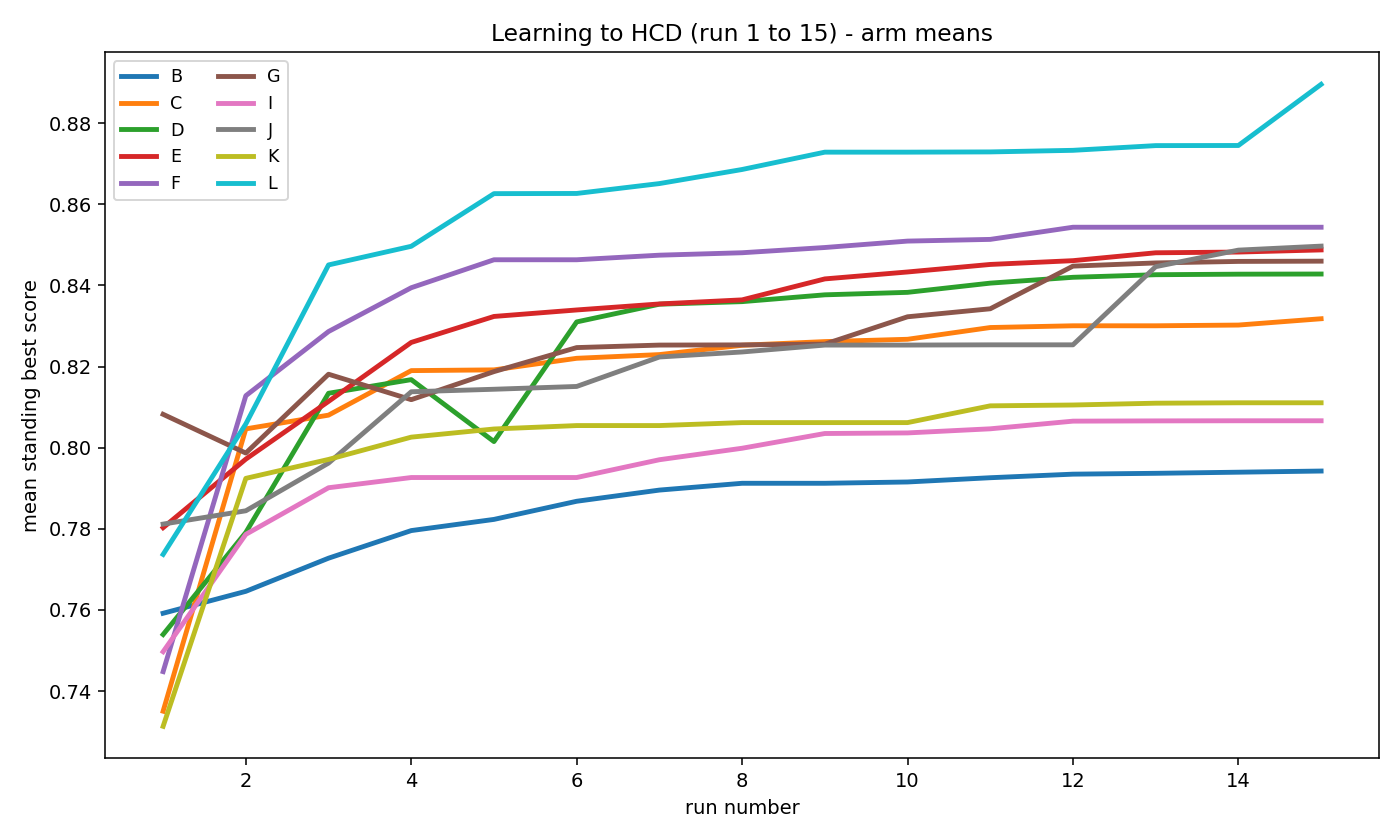}
\caption{Arm-mean learning curves (standing-best score vs.\ counted iteration, HCD window). The floor-prompt arm L dominates throughout; the full-scaffold reference B never separates from the pack despite costing the most.}
\label{fig:ablation_curves}
\end{figure}

Table~\ref{tab:nulls} gives the verdict table; Table~\ref{tab:hcd} the run-level endpoints; Figure~\ref{fig:ablation_curves} the arm-mean learning curves.

A note on pairwise power: at $n{=}10$ per arm, pairwise contrasts on the Ablation Metric detect only differences ${\geq}0.063$ (pooled per-arm $\sigma{=}0.050$, two-sided $\alpha{=}0.05$); the component-level gaps this campaign observes (0.01--0.04) run at 26--40\% power and would require 27--45 runs per arm, which is why the findings are banded against the pooled one-shot nulls (pools H and H$_G$) rather than decided pairwise. Table~\ref{tab:power} gives the per-contrast numbers; contrasts involving L's margin clear the bar, component silencing does not.

\begin{table}[H]
\centering
\caption{Pairwise power on the Ablation Metric, re-derived from per-run top scores. Minimum detectable effect at $n{=}10$ per arm: 0.063 (pooled per-arm $\sigma{=}0.050$, two-sided $\alpha{=}0.05$). Per-arm sd varies widely --- 0.025 (C) to 0.082 (B).}
\label{tab:power}
\small
\begin{tabular}{lccc}
\toprule
Contrast & Observed gap & Power at $n{=}10$ & Runs/arm for 80\% \\
\midrule
L vs B & 0.088 & 98\% & 5 \\
L vs K & 0.068 & 86\% & 9 \\
G vs K & 0.038 & 40\% & 27 \\
L vs G & 0.029 & 26\% & 45 \\
\bottomrule
\end{tabular}
\end{table}

\textbf{The loop-vs-null verdicts, by contrast, are well powered.} The verdict instrument tests each arm's ten-run mean against a Monte-Carlo reference built from the pool itself, so no arm variance is spent estimating the null: the decision edge is $+0.008$ vs.\ H ($+0.004$ vs.\ H$_G$), and 80\%-power detection is reached at gaps of 0.015--0.030 depending on the arm's dispersion --- two to four times sharper than the pairwise distinctions above.

\begin{table}[H]
\centering
\caption{Loop arms vs.\ one-shot bandit nulls. Each null pool is a set of
independent one-shot runs; the statistic is the expected best-of-15 draws
(5{,}000 Monte-Carlo simulations; all scores and verdicts are Ablation Metric). Verdict bands: \textsc{exceeds} ($p<0.05$),
\textsc{marginal} above/below ($0.05{-}0.15$, and mirrored), consistent,
\textsc{below} ($p>0.95$). Monte-Carlo quantile SE near band boundaries is
${\sim}0.005{-}0.01$, so near-band calls are unstable. Pool summary --- H
(corrected-B template, pre-ablation stack): $n{=}114$, mean 0.7644, sd 0.0736,
E[best-of-15] $=0.8463$; H$_G$ (G-package): $n{=}88$, mean 0.7802, sd 0.0659,
E[best-of-15] $=0.8455$.}
\label{tab:nulls}
\small
\begin{tabular}{lcc}
\toprule
Arm & vs.\ H null ($n{=}114$) & vs.\ H$_G$ null ($n{=}88$) \\
\midrule
\textbf{L} & \textbf{EXCEEDS} ($p{=}0.000$) & \textbf{EXCEEDS} ($p{=}0.000$) \\
F  & MARGINAL above ($p{=}0.052$) & EXCEEDS ($p{=}0.000$) \\
J  & consistent ($p{=}0.248$)     & EXCEEDS ($p{=}0.020$) \\
E  & consistent ($p{=}0.306$)     & MARGINAL above ($p{=}0.062$) \\
D  & consistent ($p{=}0.757$)     & MARGINAL below ($p{=}0.883$) \\
G  & consistent ($p{=}0.524$)     & consistent ($p{=}0.436$) \\
B, C, I, K & BELOW ($p{\geq}0.999$) & BELOW ($p{=}1.000$) \\
\bottomrule
\end{tabular}
\end{table}

\begin{table}[H]
\centering
\caption{Run-level endpoints at the common horizon (HCD$=$15 counted iterations;
carry-forward standing-best scoring). All columns are Ablation Metric scores. $n{=}10$ per arm, E$_{10}$ excluded.
L's top run reaches \textbf{0.9990}; the best of all one-shot null draws tops out
at ${\sim}0.869$ (Table~\ref{tab:nulls}). Sorted by top-run mean. Italic rows: the one-shot bandit pools, slotted by their expected best-of-15 (5{,}000-draw Monte Carlo; italic Top cell) --- the bar of Table~\ref{tab:nulls}. Arms above the italic rows clear the bar; the four below it (C, K, I, B) are the registered BELOW family. Fifteen fresh pulls cost ${\sim}$15M tokens (the ${\sim}$15 in the bandit rows' final column counts pulls, not iterations).}
\label{tab:hcd}
\small
\begin{tabular}{lcccc}
\toprule
Arm & Score at HCD & Top (mean) & Gain & Median top run \\
\midrule
\textbf{L} (floor prompt)      & \textbf{0.8896} & \textbf{0.8906} & \textbf{+0.1159} & 13.5 \\
G (no scaffold)                 & 0.8460 & 0.8612 & +0.0575 & 21.5 \\
E (no causal model)             & 0.8488 & 0.8606 & +0.0585 & 26.5 \\
F (no reconciliation)           & 0.8544 & 0.8569 & +0.0826 & 11.5 \\
J (G plus filing)               & 0.8497 & 0.8542 & +0.0685 & 23.5 \\
D (no predictions/blockers)     & 0.8428 & 0.8478 & +0.0945 & 23.0 \\
\emph{H (bandit pool: mean draw / E[best-of-15])} & 0.7644 & \emph{0.8463} & --- & ${\sim}$15 \\
\emph{H$_G$ (bandit pool: mean draw / E[best-of-15])} & 0.7802 & \emph{0.8455} & --- & ${\sim}$15 \\
C (scalar feedback)             & 0.8318 & 0.8428 & +0.0771 & 21.0 \\
K (G-family scalar feedback)    & 0.8111 & 0.8231 & +0.0736 & 15.5 \\
I (forced worksheet B)          & 0.8066 & 0.8179 & +0.0569 & 24.0 \\
B (full scaffold, reference)    & 0.7942 & 0.8027 & +0.0351 & 20.0 \\
\bottomrule
\end{tabular}
\end{table}

\begin{figure}[H]
\centering
\includegraphics[width=0.75\textwidth]{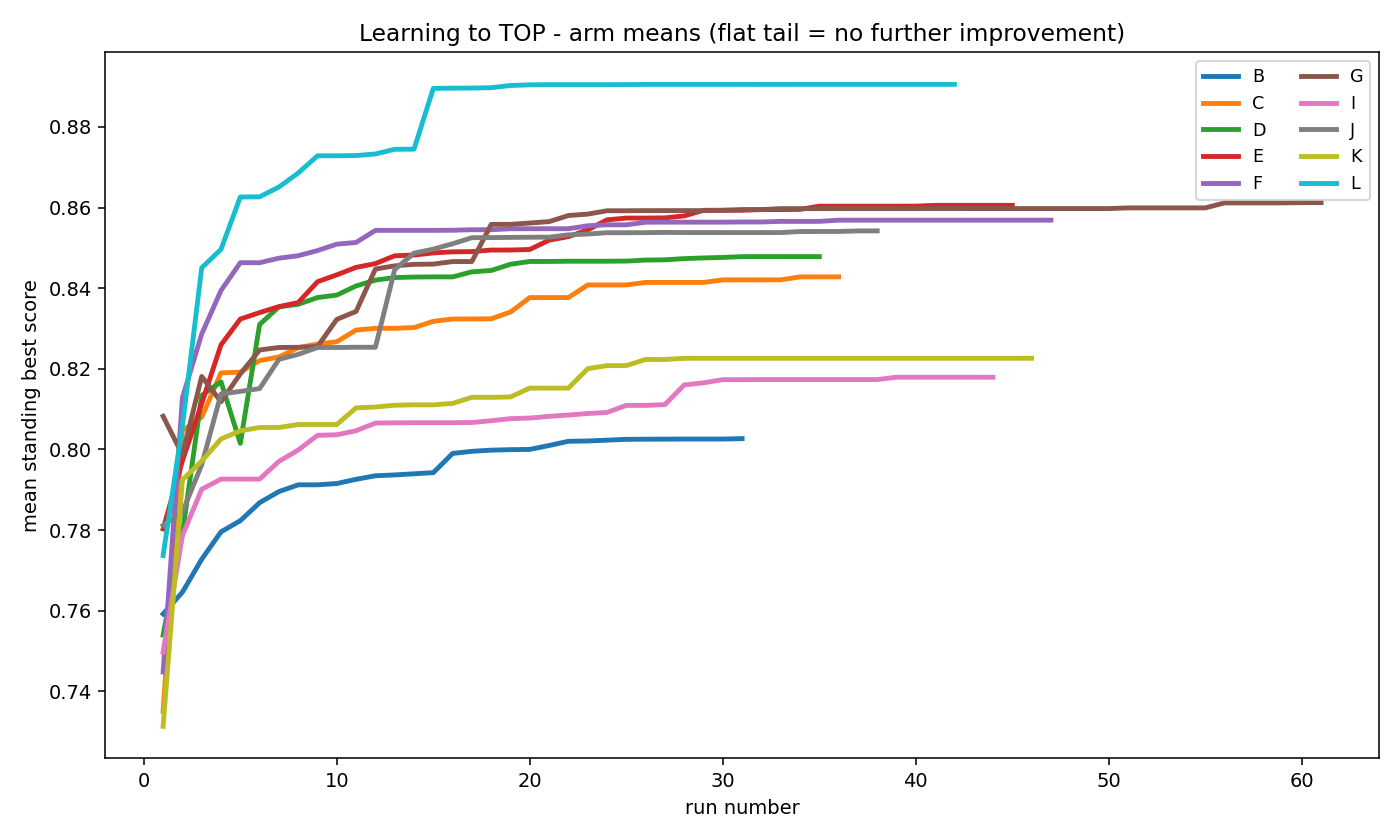}
\caption{Learning curves extended to each run's top (flat tail $=$ no further improvement). F reaches its peak soonest (median top at iteration 11.5) and L nearly as soon (13.5) at the highest level; the mid-pack arms plateau at similar levels.}
\label{fig:ablation_curves_top}
\end{figure}

\paragraph{The headline is a negative result.}
Under the G-package null, \textbf{7 of 10 arms (B, C, D, E, G, I, K) fail to beat a one-shot bandit} drawing from the same package. The full scaffold (B) is not merely indistinguishable from blind sampling --- it is \emph{below} it, decisively, in both pools, while consuming roughly $7{-}8\times$ the tokens of the pulls-only architecture (111.4M tokens/run vs.\ ${\sim}$15M for fifteen fresh pulls at ${\sim}$1.0M each). This is the scaffold tax: the deliberation apparatus costs tokens, and at this budget those tokens buy no score. The mechanism, from token forensics: the scaffold's cost is its context --- cache re-reads ${\sim}2.5\times$ the counted token total, output ${\sim}4\%$ of spend, explicit reasoning ${\sim}2.5\%$. Only L clears both pools decisively; F and J clear the H$_G$ pool but are pool-sensitive (F is only MARGINAL above the H pool at $p{=}0.052$; J is consistent with H at $p{=}0.248$). The best-of-15 statistic lives in each pool's upper tail, and the pools differ in tail shape (H: wider sd, higher max; H$_G$: narrower, higher-centered); pool-sensitive arms (F, J, E, D) should be read as ``at the one-shot level $\pm$ tail shape.'' G's loop adds nothing over G one-shots.

\paragraph{Prediction calibration}
Focusing on the prediction stage of the TLS scaffold, it asks the agent to emit pre-registered prediction intervals on what it believes the evaluation score for the algorithm will be. Of the arms where this part of the scaffold was intact (B, C, E, F, I), realized scores landed inside the agent's pre-registered prediction range only 32–61\% of the time (B 0.49, I 0.56, C 0.59, E 0.61, F 0.32), with misses predominantly below the prediction range the agent gave (B, C, E, F; I splits evenly).  The agent systematically predicted better results than its code delivered, and even accepted improvements landed outside the predicted range a third or more of the time.  

Calibration never improved over the life of each run. Making prediction quality an object of outside evaluation would likely not repair this either; instead it would create a second optimized proxy with no guaranteed coupling to artifact quality (\S\ref{sec:ablation-genomic}).

Arm D, which lacked the prediction mechanisms, lost nothing relative to the full scaffold --- the act of pre-registering an expectation did not measurably improve the results of a run.  This portion of TLS is mentioned specifically because data exists to show how it wasn't functional, whereas other parts of the ablation study show non-functionality of the part that is ablated simply due to a resulting score difference in the evaluation step.

In short, agents cannot predict outcomes with accuracy when given the information to do so.  There may be methods to improve predictive qualities in agents, but it requires something other than information about the problem, a history of past predictions, and a request.

\subsection{L: The Floor-Prompt Survivor}
\label{sec:ablation-L}

Arm L, the floor prompt, contains only the problem statement, evaluation interface, baseline, and last attempt artifact. It wins along every dimension (Table~\ref{tab:l_scorecard}); the proxy dominance does not survive transfer to the held-out benchmark (\S\ref{sec:ablation-genomic}), the economics do.

\begin{table}[H]
\centering
\caption{L: the scorecard. The floor prompt wins along every dimension --- score up \emph{and} cost down.}
\label{tab:l_scorecard}
\small
\begin{tabular}{llll}
\toprule
Ruler & L & Best alternative & Margin \\
\midrule
Ablation Metric: score at HCD (top-run mean) & \textbf{0.8896} (0.8906) & best arm & --- \\
Ablation Metric: gain over starting score & $+0.1159$ & --- & --- \\
Ablation Metric: ceiling run & \textbf{0.9990} & 0.869 (best null draw) & $+0.13$ \\
Null pools & \textbf{EXCEEDS both} & --- & $p{=}0.000$ \\
Full TF Metric & 0.8740 & 0.8757 cold draw / B 0.8774 & parity ($\leq 0.003$) \\
Value ratio (per M-token per 0.01 gain) & \textbf{2.507} & 0.289 (J) & ${\sim}9\times$ \\
\bottomrule
\end{tabular}
\end{table}

\paragraph{No state accumulation at all.}
One worry was that even a minimal loop would pile up context as it went --- each turn reading more of the run's history than the last, the growth pattern compression theory predicts. We recovered the trajectories to check (method below), and the worry never materializes. L's context runs between 0.25 and 0.75M tokens per iteration the whole way, and the early iterations are actually the \emph{larger} ones: 0.52M on average over the first three, versus 0.32M over the last three, for a median run-level slope of $-0.015$M per iteration (Wilcoxon $p{=}0.0195$, $n{=}10$). K behaves differently but tells the same story --- its context is spiky, with isolated 12--13M single-search attempts at iterations 3--4 and 8, yet never trends upward (slope $-0.601$M/iter, $p{=}0.0117$, $n{=}9$). Neither arm shows the growth compression theory worried about (Table~\ref{tab:context_traj}). Whatever carried these arms, it was not a growing memory.

\begin{table}[H]
\centering
\caption{Per-iteration context trajectories for the two minimal-context arms
(method: history timestamps + transcript mtimes; K and L only). Neither arm
shows context growth: L is flat-to-shrinking throughout; K is spiky (isolated
heavy single-search attempts) but also non-growing.}
\label{tab:context_traj}
\small
\begin{tabular}{lcc}
\toprule
 & \textbf{L} (floor prompt) & \textbf{K} (G-family scalar) \\
\midrule
Per-iteration tokens          & 0.3--0.75M, tiny throughout & spiky \\
Early vs.\ late iterations    & iters 1--3 mean 0.52M       & spikes at iters 3--4 and 8 \\
                              & vs.\ iters 13--15 mean 0.32M & (12.6M, 12.9M) \\
Run-level slope (median)      & $-0.015$M/iter              & $-0.601$M/iter \\
Slope test (Wilcoxon)         & $p=0.0195$ ($n{=}10$)       & $p=0.0117$ ($n{=}9$) \\
\bottomrule
\end{tabular}
\end{table}

\subsection{Held-Out Genomic Benchmark: Loop--Bandit Parity}
\label{sec:ablation-genomic}

One facet of the ablation study was that the evaluation was on a proxy of 15 synthetic TFs to the larger evaluation used to measure motif discovery that led to DALE (53 TF genomic negatives). As such, it was possible (and observed) that agent runs could max out the score on the proxy.  We wanted to see if these scores on the proxy transferred to higher scores on the full benchmark.

To measure what survived, all
256 retained fleet artifacts (per-run quantile snapshots) plus the two
one-shot pools were scored on the 53-TF genomic-negatives benchmark --- the
\textbf{Full TF Metric} throughout.

\textbf{Parity.} The fleet's best artifact is B's 0.8774; L's best is 0.8740;
the best cold draw from the one-shot pools reaches 0.8757 ($n{=}197$ scored
H-family artifacts; all Full TF Metric). B took much longer to reach that point than L, and the fact that H arrived there at all again underscores that LLM-agents work well when sampled.

\textbf{The proxy decouples near its ceiling.} Within run G\_5 itself, the
benchmark tracks the wrong way as the proxy optimizes: 0.816 (baseline-era
artifact) $\rightarrow$ 0.844 (mid-run) $\rightarrow$ \textbf{0.699} (the
late, eval-optimized artifact) --- the steepest within-lineage collapse in the
corpus. L's mid-run region, by contrast, climbs 0.859 $\rightarrow$ 0.874.

\textbf{Spread.} Individual one-shot draws range from 0.62 to 0.876 on the
benchmark.

\begin{figure}[H]
\centering
\includegraphics[width=\textwidth]{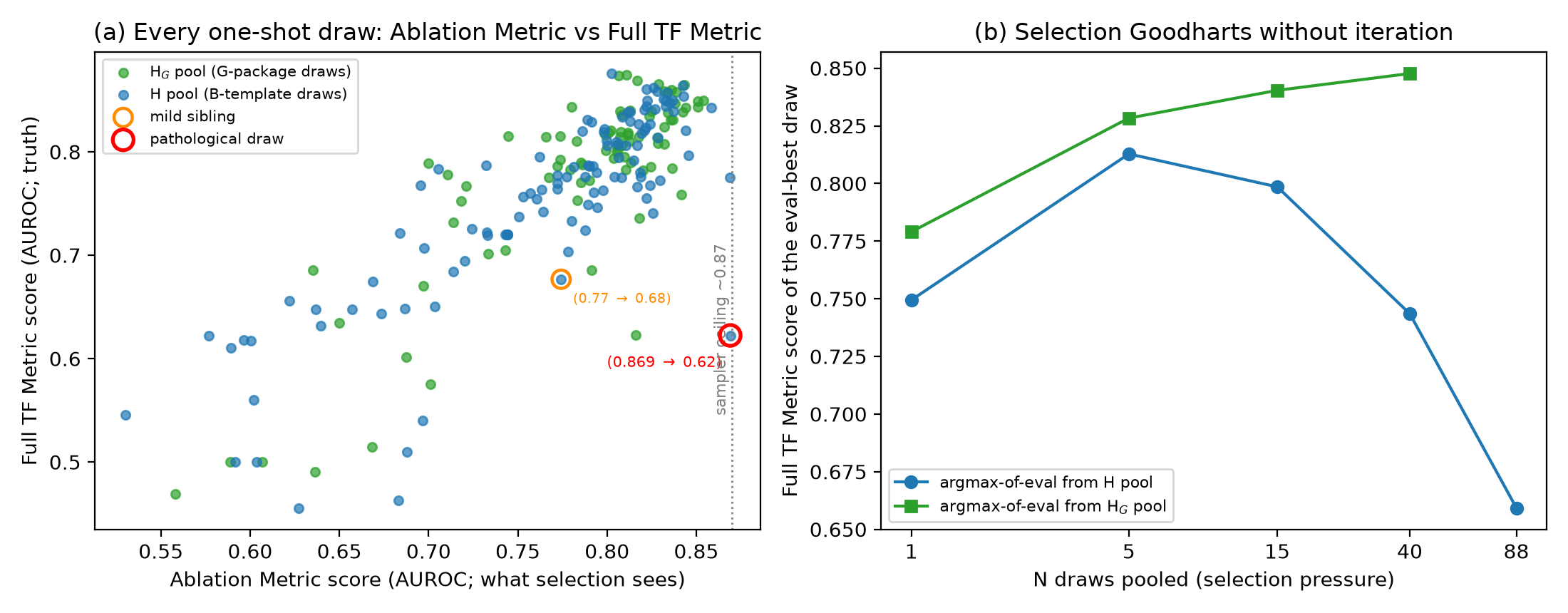}
\caption{Selection Goodharts without iteration. (a)~Every scored one-shot draw, Ablation Metric score vs.\ Full TF Metric score (114 H draws, 83 H$_G$ draws): proxy and truth track through the bulk, and the coupling breaks past the sampler ceiling (${\sim}0.87$), where the H pool's eval-best draw (red ring) is pathological. (b)~Benchmark score of the eval-argmax draw when pooling $N$ draws (5{,}000 Monte-Carlo draws per point): the H curve declines as $N$ grows --- selection converges on the pathological draw --- while the G-package pool, free of pathological members, saturates monotonically.}
\label{fig:goodhart_selection}
\end{figure}

\textbf{Selection Goodharts without iteration.} It would be convenient to blame the loop --- to imagine that iterating on the proxy is where the corruption enters. The one-shot pools let us test that directly, and they say otherwise: selection alone reproduces the failure with no iteration at all (Figure~\ref{fig:goodhart_selection}).

\subsection{The Evolve Family: Sampling and Selection Reach the Same Frontier}
\label{sec:ablation-armn}

\begin{figure}[H]
\centering
\includegraphics[width=0.9\textwidth]{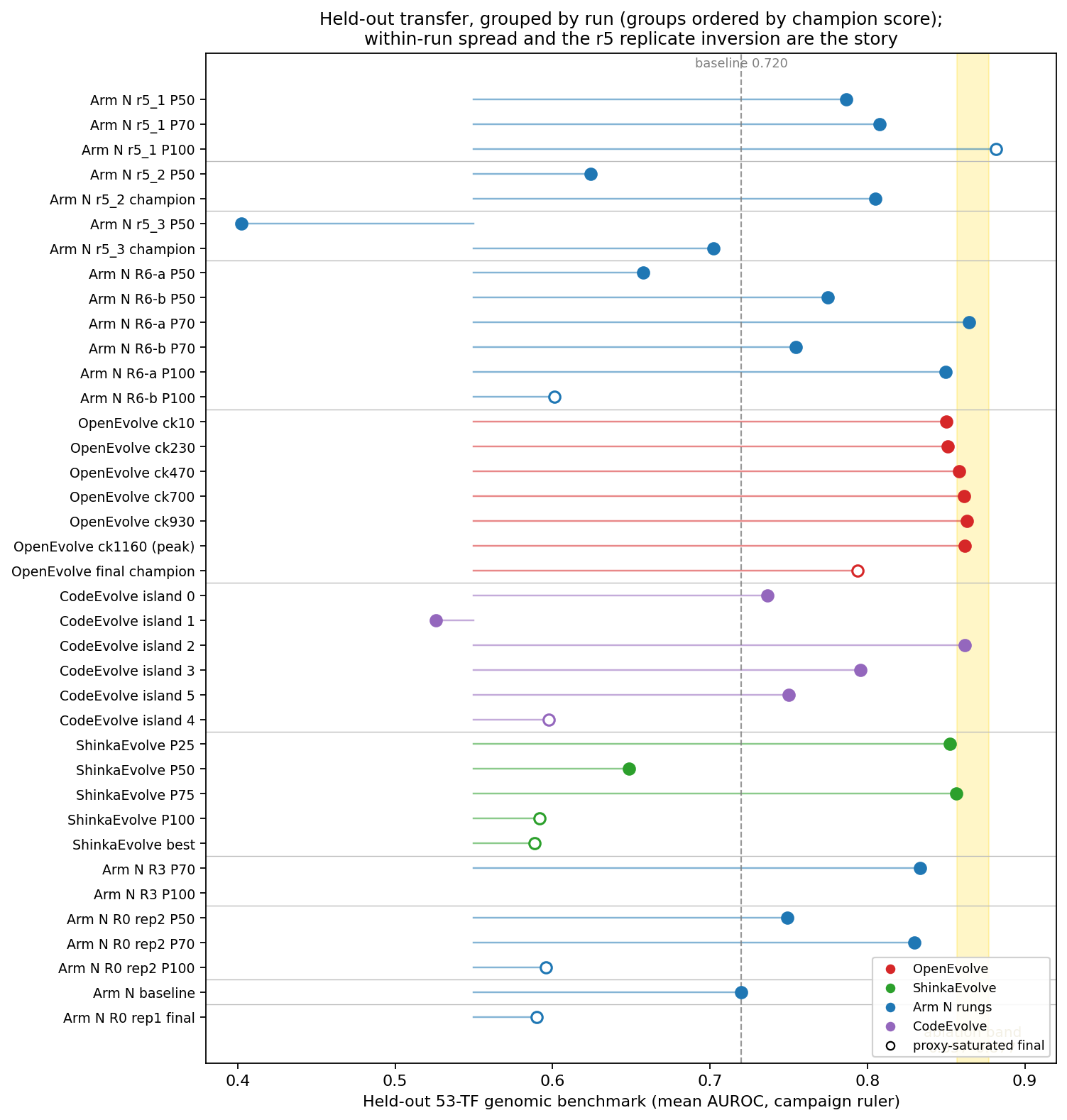}
\caption{Arm N and the Evolve family on the held-out 53-TF genomic benchmark (Full TF Metric): each artifact's mean AUROC against the ablation band. Every family places mid-run artifacts inside the band; open circles mark each run's proxy-saturated final, which falls below it. r3\_1's 1.0197 artifact (which returns no candidates on real data) is omitted.}
\label{fig:armn_regions}
\end{figure}

A third architecture family ran the same task alongside the campaign: three open code-evolution frameworks --- OpenEvolve (an open-source implementation of the AlphaEvolve architecture), ShinkaEvolve, and CodeEvolve --- plus Arm N, a scientist-proxy ladder of CodeEvolve-engine runs varying the agentic layer over the same engine:
\begin{itemize}
\item \textbf{r0} --- stock replicates, no agentic layer;
\item \textbf{r3\_1} --- island cabinets;
\item \textbf{r5} --- framing-only;
\item \textbf{filing2/3} --- state-plus-framing, i.e.\ the J-arm concept on CodeEvolve.
\end{itemize}
Run designations for every family are tabulated in the Run Registry (Appendix~F). These frameworks embody the opposite bet from the scaffold: no scientific-method apparatus at all, just populations, mutation, and external selection. Their histories were recovered and hash-joined to artifact code, and 36 artifacts spanning every progression were scored on the 53-TF benchmark (Figure~\ref{fig:armn_regions}).

\begin{table}[H]
\centering
\caption{The Evolve family on the held-out 53-TF benchmark.}
\label{tab:armn_frontier}
\small
\begin{tabular}{llll}
\toprule
Member & Held-out AUROC & Markov-1 & Note \\
\midrule
OpenEvolve & 0.862 & 0.882 & enters ablation band by eval 10 (0.850), peak by eval ${\sim}$1000 \\
ShinkaEvolve & 0.852--0.856 & --- & mid-run artifacts \\
CodeEvolve & 0.862 & --- & --- \\
Scientist-proxy rungs & 0.830--0.865 & --- & r0 through filing2/3 \\
\bottomrule
\end{tabular}
\end{table}

\textbf{The best held-out score in the entire study comes from this family --- as its best draw.} Run r5\_1 --- the framing-only rung, proxy peak just 0.8364 --- scores \textbf{0.8819} on the Full TF Metric and \textbf{0.8964} on STREME's Benchmark, exceeding every fleet artifact and the pre-ablation binary itself. 

For context against the case-study's own tools, Table~\ref{tab:abl_gen_ref} places the ablation-era artifacts --- including r5\_1 --- under the case study's native Markov-1 scoring (STREME's Benchmark); the Full TF Metric parity analysis of \S\ref{sec:ablation-genomic} remains the study's own measurement.

\begin{table}[H]
\centering
\caption{Genomic negatives benchmark (53 TFs, STREME's Benchmark --- Markov-1 scoring), replicated
from the case study (Table~\ref{tab:main_gen}) with the ablation artifacts
restored. Upper block: reference-tier tools. Lower block: this study's
artifacts --- the discovered algorithm and its ablation-era relatives ---
scored under the identical eval.}
\label{tab:abl_gen_ref}
\small
\begin{tabular}{lrr}
\toprule
Tool & Mean AUROC & Time/TF \\
\midrule
DALE             & 0.891 & 0.3\,s \\
MEME 5.5.5       & 0.881 & $\sim$90\,s \\
proto-DALE       & 0.874 & 1.4\,s \\
STREME 5.5.5     & 0.863 & 4.4\,s \\
\midrule
\textbf{r5\_1 (Arm N framing rung)} & \textbf{0.8964} & single candidate \\
B\_1\_P70 (ablation arm B)         & 0.8896 & 12 candidates \\
\bottomrule
\end{tabular}
\end{table}

One caveat travels with these numbers: artifacts from the final, proxy-saturated phase of each run (the ${\sim}1.0$-scored champions; open circles in Figure~\ref{fig:armn_regions}) transfer poorly --- 0.58--0.60, or no candidates at all on real data --- while mid-run artifacts do not --- the three framing-rung replicates extend the pattern below saturation, inverting transfer against proxy peak across the 0.84--0.92 range; \S\ref{sec:ablation-genomic} shows the same pattern inside the fleet. Transfer damage grows with the intensity of proxy optimization, which is why late-run proxy surges are discounted in the selection rules above.

\section{Discussion}
\label{sec:discussion}

\subsection{What 'The Little Scientist' Discovered}

Despite the later discovery that the TLS framework is largely not the causal mechanism behind LLM based algorithmic search, the process nevertheless uncovered a novel algorithm for DNA motif discovery.

The search produced an algorithm that combines established and novel design choices: k-mer enumeration as a seeding strategy and EM as a refinement mechanism (both well-known in the literature), assembled new strategies: dual seeding that scores candidates by both abundance and consistency, Pareto-front ranking that preserves diverse width strategies, and width--aware scoring that penalizes uninformative positions. On the shuffled-negative benchmark the artifact outscores STREME (0.842 vs.\ 0.803).

\subsection{What 'The Little Scientist' Ablation Study Discovered}

What the search statistics describe is more specific than first assumed. Under the full framework, 17 of $\sim$64 evaluated iterations produced improvements (26.6\%). We initially read these as the signature of hypothesis-driven reasoning, with failures constraining the design space, as in human algorithm development. The controlled ablation campaign revised that reading: one-shot draws from the same model, with no hypotheses, no memory, and no reconciliation, reproduce the same improvement economics and the same ceiling (\S\ref{sec:ablation-verdicts}).  Sampling with environmental level selection is the reason this process worked at all.

The ablations prove that LLM-agents do not make use of an imposed process (TLS) to build up a model of problem sets over time.  They do not use such a model to iterate further solutions based on refined hypotheses.  In fact, although they seem to be able to conform to the methods imposed on them to follow, the scores from the evaluations show that whereas such scientific processes might help a person, they hinder or slow an LLM in reaching positive results.  Agents bring their knowledge to bear mostly in the form of innate training, sampling and recombining from that storage.  The ablation arm that had the least ceremony (L), and incidentally which is most in line with something like AlphaEvolve, did best.

\subsection{Implications for Agentic Science}

The practical implications to the above:
\begin{itemize}
\item \textbf{Spend on draws and selection, not deliberation:} the minimal loop matched or beat every scaffolded configuration, and the one-shot nulls matched the loops --- independent lineages outperform deeper reasoning per token (\S\ref{sec:ablation-verdicts}). The full token economics behind this are quantified in the appendix: value ratios in Table~\ref{tab:value_ratio}, with the per-arm economics figures (gain per token, cost per point, per-attempt breakdown) alongside.

\item \textbf{Feedback matters only where the loop beats sampling:} the null draws never touch a feedback channel (no diagnostics, not even a scalar score), yet best-of-15 sampling matches or beats 7 of 10 loop-based ablations (Table~\ref{tab:nulls}). Only L clears both pools decisively, with F marginal against H ($p{=}0.052$). Below the outcome bar, feedback richness still shows at the mechanism level, consistent with arguments for dense epistemic signals \citep{he2026agentic}: on the same package, swapping full per-instance diagnostics for a scalar digest is worth $0.038$ in top-run mean (Ablation Metric; G 0.8612 vs K 0.8231; bootstrap 95\% CI $-0.081$ to $-0.006$).

\item \textbf{The winning ablation informs why code-evolve architectures work:} we ran three open code-evolution frameworks with no scientific-method apparatus, and they reached the same held-out frontier as everything else (\S\ref{sec:ablation-armn}); the more the mechanisms that were internal to TLS were externalized, and the more that the use of the LLM was minimized to the essentials to what it needed for the problem, the better it performed, in line with Ablation L, and in line with typical code-evolve architectures.

\end{itemize}

\subsection{Mining, Not Building}
\label{sec:discussion-mining}

Across every instrument in this study --- the controlled ablations, the
one-shot bandit nulls, the held-out genomic benchmark, the ${\sim}300$-run
pre-ablation lineage, and the three-framework Evolve family --- one boundary
held: \textbf{no process reliably exceeded what its prior contained.}
Single artifacts grazed or passed the prior's best cold draw, but no
exceedance replicated, and no process mean cleared the prior's frontier
(derivations and sources: Appendix~G). Loops proved valuable over bandits on the proxy eval's progression (L decisively, F marginally), but the harder the proxy was chased, the more the full eval was overfit, such that bandit pulls achieved similar results on the Full TF Metric (\S\ref{sec:ablation-genomic}). The value of looping is demonstrable; what this design cannot separate is loop progress from draw luck, because the two converge on the same held-out frontier. Further experiments of this nature on a different problem set might show loops to have value in both proxy and held-out evaluations.

\textbf{What loops delivered was reliable extraction, not
frontier-pushing.} The prior's best cold draw (0.8757 on the Full TF
Metric, at a mid-pack proxy of 0.8023) was at parity with L's best
artifact (0.8740), and the Evolve-family
means at 0.85--0.86 are similarly placed. L's contribution is reliability,
not reach: all ten runs at or above the bandit bar, top-run mean $+0.044$
over E[best-of-15] on the Ablation Metric ($p{\approx}0.000$).

\textbf{Loops exceed their priors routinely on the Ablation Metric, and that is the channel that overfits the Full TF Metric.} L's best run reaches
0.999 and G\_5's 1.0193 against cold-draw maxima of 0.869 (B-template
pool) and 0.854 (G-package pool); yet inside run G\_5 itself, the Full TF
Metric falls as the proxy climbs (0.816 $\rightarrow$ 0.844 $\rightarrow$
0.699; \S\ref{sec:ablation-genomic}).

What the processes did deliver: prior plus selection beat a decades-tuned
incumbent (STREME) by recombination. Mining the weights works.

\subsection{Future Directions}
    
    \textbf{Benchmark dependency.} The framework requires a well-defined benchmark and ground-truth evaluation data.  More effort directed toward problem formalization (encoding scientific questions as benchmarkable tasks with structured data) would unlock autonomous discovery across a wider range of domains.

\vspace{1em}
\noindent\textbf{Data availability.}

\emph{Motif discovery.} All benchmark data used in this study is publicly available. ChIP-seq peak sequences for 132 ENCODE K562 TFs were obtained from the ENCODE Consortium \citep{encode2012} (\url{https://www.encodeproject.org}). Peak regions were processed into 100\,bp centered windows around summits. JASPAR 2024 \citep{castro2024jaspar} motif profiles (\url{https://jaspar.elixir.org}) were used as reference motifs for TOMTOM comparison. Shuffled and genomic negative sequences were generated from the same assemblies using standard protocols. No custom or proprietary datasets were generated.

\vspace{1em}
\noindent\textbf{Code availability.}

\emph{DALE.} The discovered algorithm is distributed as a statically-linked binary and Python interface, along with the full C source, benchmark scripts, and a reproducible Jupyter notebook, at \url{https://github.com/travis42/little-scientist-dale}.

\emph{Delta V.} The discovered ensemble strategy (a Python function), the evaluation pipeline, and the full search history (all archived iterations) are available at \url{https://github.com/travis42/little-scientist-delta-v}.

\emph{Framework.} The Little Scientist framework --- including agent prompts, evaluation environment code, the Kuhn agent, and the cron-driven orchestration --- is available at \url{https://github.com/travis42/little-scientist-dale}. The framework uses the OpenClaw agent platform (\url{https://github.com/openclaw/openclaw}) for LLM invocation, scheduling, and sandboxed execution. No fine-tuning or training of language models was performed; API access for inference is sufficient. The LLM used was GLM-4.7 (Z.ai), accessed via a coding plan subscription.

\vspace{1em}
\noindent\textbf{Acknowledgements:} The author thanks the ENCODE Consortium for providing open-access ChIP-seq data, the JASPAR database for motif reference data, and the MEME Suite development team --- Timothy Bailey, Charles Grant, and William Noble --- whose tools (STREME, MEME, TOMTOM) were used extensively in this study and whose work has advanced the field of computational motif discovery over three decades. Portions of this manuscript's prose were drafted with assistance from z.ai's GLM 5.1 (the writing assistant; the experiments themselves ran the GLM-4.7 agent of \S\ref{sec:ablation-campaign}), under the author's direction; the author reviewed, edited, and takes full responsibility for all content.

\vspace{1em}
\noindent\textbf{Funding:} This work received no external funding.

\bibliographystyle{plainnat}
\bibliography{bibliography}

\clearpage
\appendix
\renewcommand{\thesection}{\Alph{section}}
\section*{Supplementary Material}

\subsection*{Section A: Per-TF Results}

Per-TF AUROC values for all 132 TFs are provided in the accompanying benchmark data files (\texttt{benchmark\_data/shuffled\_negatives\_132tf.csv} for the Case-Study Benchmark, and \texttt{benchmark\_data/genomic\_negatives\_132tf.csv} for the genomic-negatives scoring, whose 53-TF held-out subset is the Full TF Metric of \S\ref{sec:ablation-genomic}).

\subsection*{Section B: Compilation Fairness}

STREME prebuilt binary vs.\ source-compiled (\texttt{-O2}) across 23 TFs: ratio 1.00$\times$ (identical performance). Both compiled with GCC on Ubuntu 24.04.

\subsection*{Section C: Reproducibility}

All benchmark data, the DALE binary, and a Jupyter notebook reproducing all motif discovery figures and statistics are available at \url{https://github.com/travis42/little-scientist-dale}. The repository includes reproducible benchmark scripts and is released under the Apache 2.0 license.

\section*{Section E: Robustness --- Excluding Optimization Transcription Factors}

During development, DALE's iterative optimization loop received feedback on a fixed set of
15 ChIP-seq TFs, selected adversarially as the algorithm's weakest
cases relative to STREME at the time the set was constructed (five diagnostic TFs where
DALE trailed STREME, five regression-guard TFs where DALE led, five stability TFs).
All 15 TFs are contained in the 132-TF ENCODE K562 evaluation set. This appendix reports
all headline comparisons recomputed on the \emph{disjoint} subsets containing none of the
15 optimization TFs, using the same per-TF measurements as the main results. No algorithms
were re-run; this is a re-analysis of the existing per-TF data under exclusion.

\begin{table}[H]
\centering
\caption{DALE vs.\ STREME with all 15 optimization TFs excluded. Win/loss counts are
paired per-TF comparisons; $p$-values from Wilcoxon signed-rank tests. All rows
are same-metric comparisons from the packaged binary's Markov-1 scoring (the
disjoint rows exclude the 15 TFs the search optimized on); study artifacts under
this metric are reported in Table~\ref{tab:main_gen}.}
\begin{tabular}{lccccc}
\toprule
Benchmark & $n$ & DALE & STREME & $\Delta$ & W/L \\
\midrule
Shuffled neg.\ (full)      & 132 & 0.842 & 0.803 & $+0.039$ & 89/43 \\
Shuffled neg.\ (disjoint)  & 117 & 0.837 & 0.797 & $+0.040$ & 79/38 \\
Genomic neg.\ (full)       & 53  & 0.891 & 0.863 & $+0.028$ & 36/17 \\
Genomic neg.\ (disjoint)   & 38  & 0.905 & 0.881 & $+0.023$ & 26/12 \\
Optimization TFs (the 15)  & 15  & 0.881 & 0.852 & $+0.029$ & 10/5 \\
\bottomrule
\end{tabular}
\end{table}

Three observations:

\textbf{(1) The headline result is unchanged on training-disjoint data.} On the 117-TF
disjoint shuffled set, DALE's margin is $+0.040$ AUROC ($p < 10^{-20}$, 79 wins / 38
losses) --- marginally larger than the full-set margin. On the 38-TF disjoint genomic
subset, the margin is $+0.023$ ($p \approx 8\times10^{-8}$). DALE's advantage over
STREME is not carried by the TFs its optimization loop received feedback on.

\textbf{(2) Test-set contact inflation is small and now bounded.} DALE's absolute AUROC
on the 15 optimization TFs (0.881 shuffled mean) exceeds its disjoint-set mean (0.837);
because those 15 TFs constitute 11\% of the 132-TF set, the aggregate inflation
attributable to optimization-set inclusion is $\approx 0.005$ AUROC (0.842 reported vs.
0.837 disjoint). All headline margins exceed this by roughly an order of magnitude.

\textbf{(3) Optimization-set selection was adversarial, not favorable.} The 15 TFs were
chosen where DALE performed \emph{worst}, i.e., the optimization loop trained on failures.
This selection direction cannot inflate the disjoint-set result by construction: no
information from the 117 disjoint TFs (or STREME's behavior on them) was available to the
optimization process.

We report this analysis because the optimization set overlaps the evaluation benchmark ---
a condition that, left unexamined, could confound tool-comparison claims. The exclusion
re-analysis bounds the effect of that overlap and is, to our knowledge, a check that is
rarely reported for methods tuned iteratively against public benchmarks.

\section*{Section F: Run Registry}

Internal designations used throughout this paper, mapped to their designs.

\subsection*{Controlled-ablation arms (the fleet)}
\begin{tabular}{lp{0.62\textwidth}}
B & full scaffold (reference): the pre-ablation stack \\
C & scalar feedback only: per-instance diagnostics replaced by an aggregate score \\
D & no predictions/blockers: pre-evaluation reasoning removed \\
E & no causal model: persistent mechanism memory removed \\
F & no reconciliation: prediction post-mortems removed \\
G & no scaffold: entire epistemic layer deleted; generate-and-test with scalar feedback \\
H & one-shot null: independent pulls from the corrected-B template (H$_G$: G-template pulls) \\
I & forced worksheet: full scaffold with enforced worksheet format (exploratory, retroactively disclosed) \\
J & G plus filing: G's free-form loop plus artifact files as pure state records \\
K & G-family with scalar feedback \\
L & floor prompt: problem statement, eval interface, baseline, last attempt \\
\end{tabular}

Run subscripts index replicates (B$_1$--B$_5$, G$_5$, J$_5$, \dots);
quantile suffixes mark artifacts sampled by run position (P0 first valid
eval, P50/P70 median/late, P100 best proxy artifact) --- e.g.\ B$_1$\_P70
is the 70th-percentile artifact of replicate~1 of arm~B.

\subsection*{Arm N: the CodeEvolve-engine proxy ladder}
Six islands per run, CodeEvolve engine throughout; rungs differ only in the
LLM-proxy layer over the same engine.
\begin{tabular}{llp{0.5\textwidth}}
r0\_rep1, r0\_rep2 & stock & no proxy layer (direct API) --- the CodeEvolve baseline \\
armn\_r3\_1 & R3 & island cabinets: per-island hypothesis/plan/blocker state \\
r5\_1, r5\_2, r5\_3 & R5 & framing-only: paradigm framing text, no persistent state (three replicates; r5\_2/r5\_3 added 2026-09-23) \\
filing2, filing3 & R6 & state + framing: the J-arm concept (filing/state records) ported onto CodeEvolve \\
\end{tabular}

\subsection*{Evolve-family frameworks}
OpenEvolve (open-source AlphaEvolve architecture): the letitrip lineage;
checkpoint numbers (ck10--ck1780) are eval counts. ShinkaEvolve run~3
(SHINKA3): 1{,}179 evals; P-suffixes as above. CodeEvolve islands
(CDEV): per-island terminal bests of the main run.

The \emph{ablation band} is the 0.857--0.877 span in which the best
held-out artifact of every controlled-ablation arm landed on the 53-TF
benchmark; every architecture family tested reaches it.

\section*{Arm-by-Arm Conclusions}
\label{app:arm-conclusions}

The per-arm reading behind \S\ref{sec:ablation-verdicts}, stated with its guards.

\begin{tabular}{lp{0.30\textwidth}p{0.42\textwidth}}
\textbf{Arm} & \textbf{Result} & \textbf{Defensible conclusion} \\
B & below blind sampling at $7{-}8\times$ tokens & CAN follow an elaborate epistemic protocol faithfully; CANNOT convert deliberation into search advantage at this budget. \\
C & parity with B & A scalar score plus the standing best suffices for targeting; per-instance diagnostics bought nothing. Guard: the floor result (0.664 stall with no feedback) keeps the claim flat-above-the-floor, not never-matters. \\
D & parity with B & No discipline effect from pre-registering expectations; intervals systematically wrong despite the closed in-context feedback loop. Untenable at the mechanism's design point. \\
E & no measurable return (pool-sensitive) & Models rebuild working context from minimal files each turn; accumulated mechanism notes paid nothing. Not read as proof of uselessness. \\
F & clears the H$_G$ pool, marginal over H ($p{=}0.052$); fastest to peak (median iteration 11.5); return-stuffing 0/38 vs.\ 13/61 in L & Post-mortems are behaviorally real, competitively inert: reconciliation changed what the diffs contained without changing what they achieved. \\
G & G$_5$ = study proxy maximum (1.019); loop adds nothing over G one-shots & A strong model, a scalar signal, and keep-if-better is the engine: frontier-level proposal generation under bare selection. \\
H & mean 0.7644; per-run learning curves reproduce from draws alone & A rising within-run score curve is not by itself evidence of within-run learning, because draws plus the keep-best rule reproduce the same shape. Guard: null scoped to the same package --- no claim about other packages; and L's decisive margin over both pools shows the loop itself adds value beyond fresh draws. \\
I & decisively below the null; calibration misses split evenly (the only non-optimistic arm) & Enforced epistemic format produces compliance without even the optimism signature of voluntary prediction, at cost to score. The calibration split is an observation, not a claim. \\
J & ceiling 1.005; MIXED on its registered discriminator; pool-sensitive & State-as-records without selection pressure is a passenger --- independently confirmed by Arm~N's R6 runs. \\
K & decisively below both nulls & Degrading the feedback channel to scalar-only dropped the loop from break-even to below-sampling: same package as G, whose loop matched its null. Feedback richness buys loop viability, not frontier --- G's endpoints sit ${\sim}0.04$ above K's (point estimate, $n{=}10$). \\
L & the only arm decisively above both null pools; ceiling 0.999; order-of-magnitude value ratio & Four lines of context --- problem, eval interface, baseline, last attempt --- suffice. Everything beyond is overhead with a price tag. \\
\end{tabular}

\begin{figure}[H]
\centering
\includegraphics[width=0.75\textwidth]{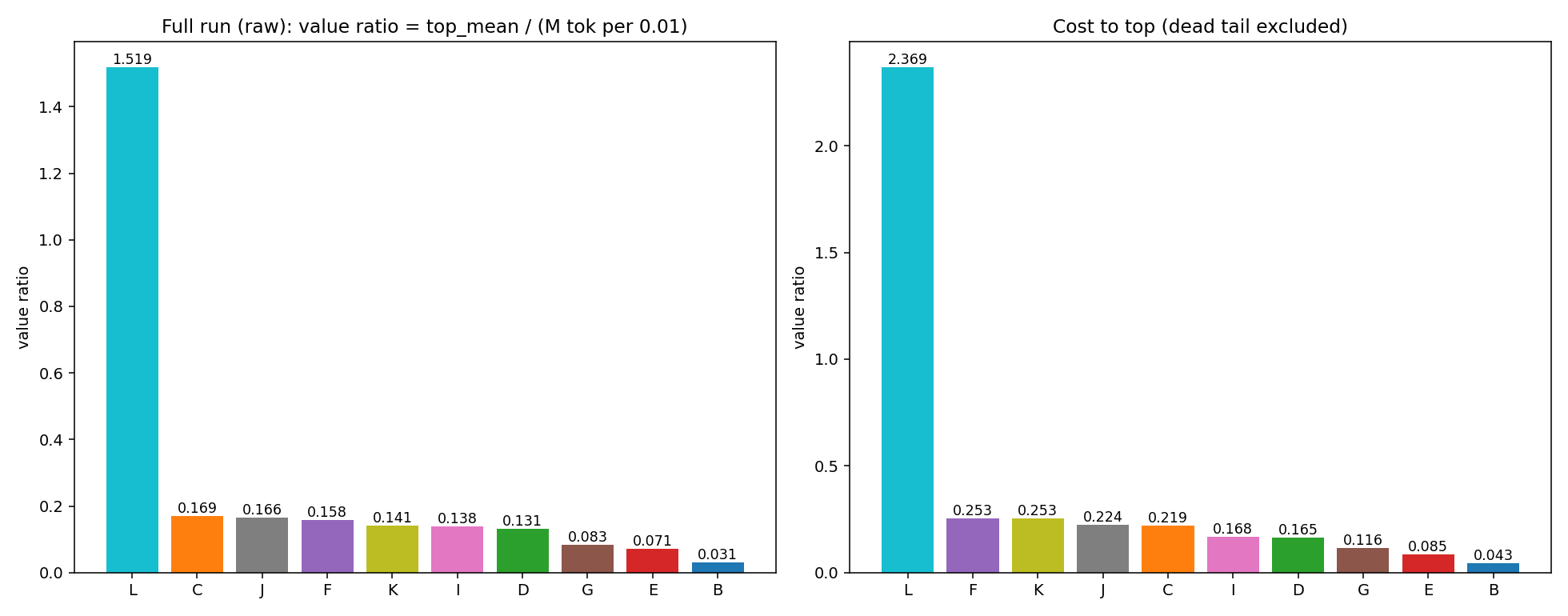}
\caption{Value ratio over the full run (top score per M-token per 0.01 gain).}
\label{fig:ablation_value_ratio_full}
\end{figure}

\paragraph{Economics, full-run view.}
The HCD-window value-ratio ranking of Table~\ref{tab:value_ratio} is
unchanged when the whole trajectory is charged
(Figure~\ref{fig:ablation_value_ratio_full}).

\begin{figure}[H]
\centering
\includegraphics[width=0.75\textwidth]{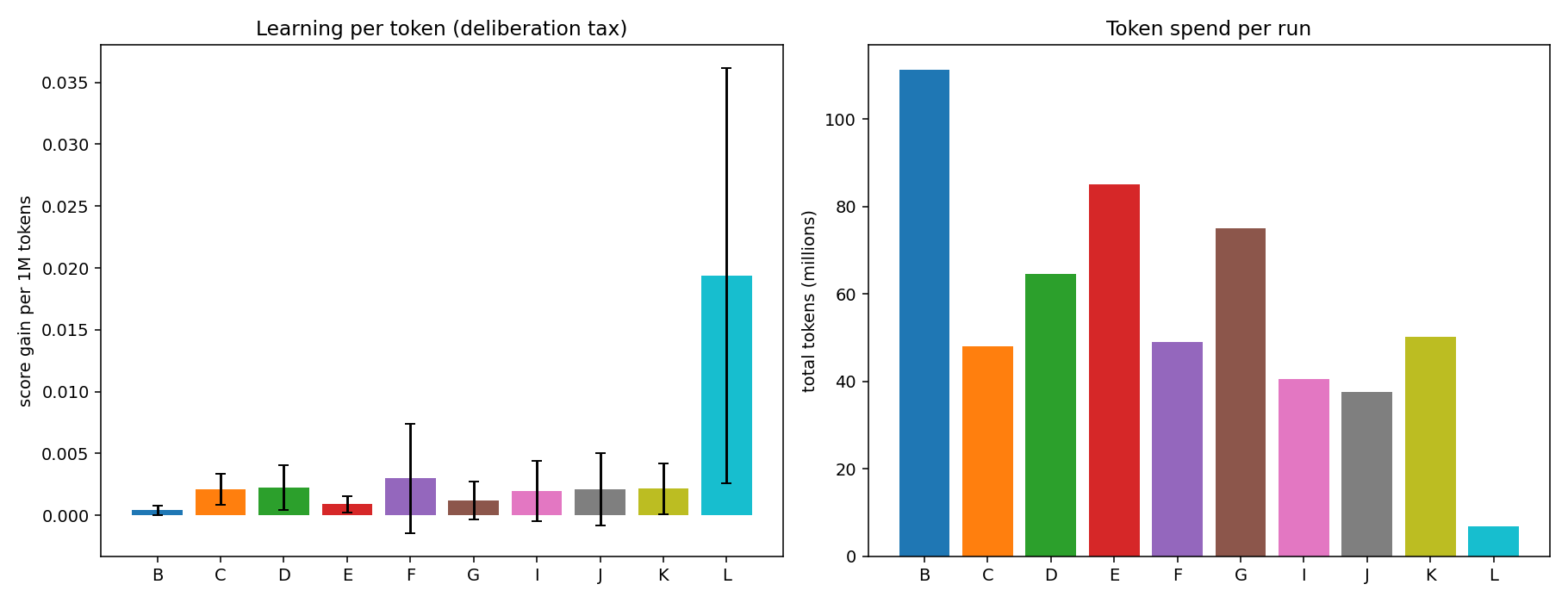}
\caption{Gain per token by arm (value composite). L's dot sits an order of magnitude above the field.}
\label{fig:ablation_gain_per_token}
\end{figure}

\begin{figure}[H]
\centering
\includegraphics[width=0.75\textwidth]{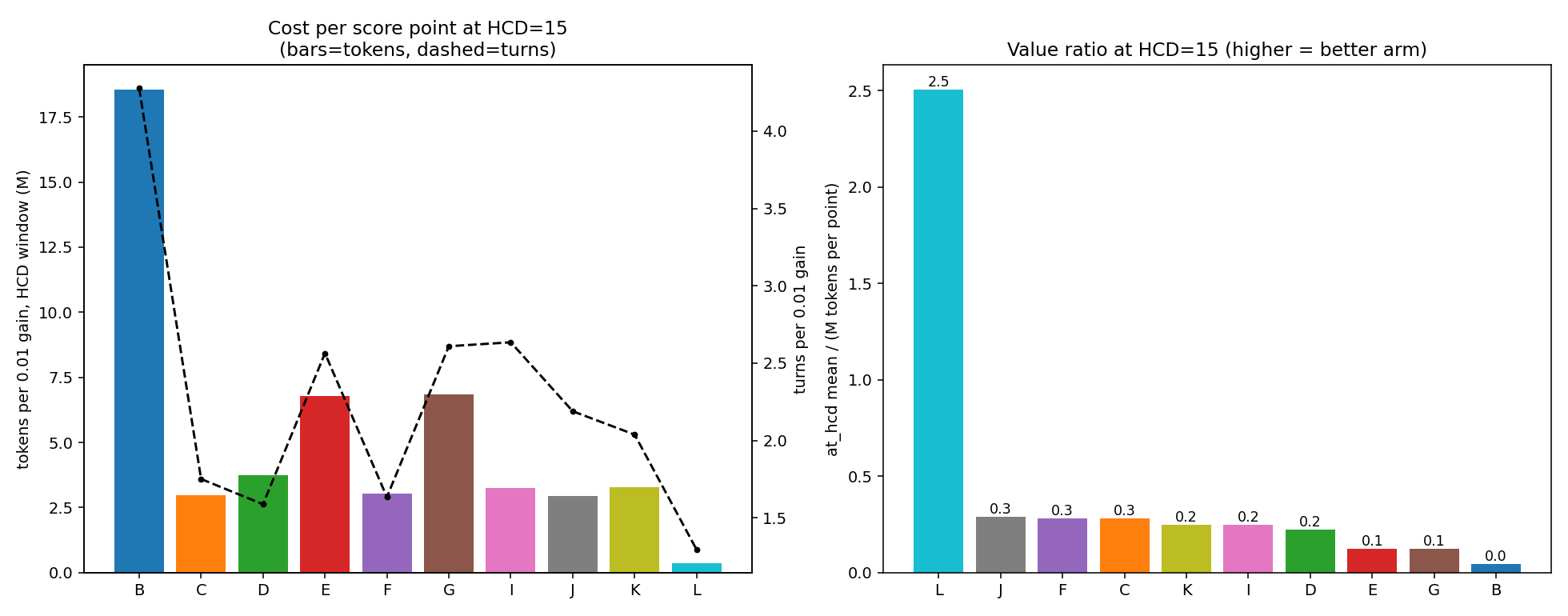}
\caption{Cost per score point (tokens per 0.01 gain, HCD window). B pays ${\sim}50\times$ L's rate.}
\label{fig:ablation_cost_per_point}
\end{figure}

\begin{figure}[H]
\centering
\includegraphics[width=0.75\textwidth]{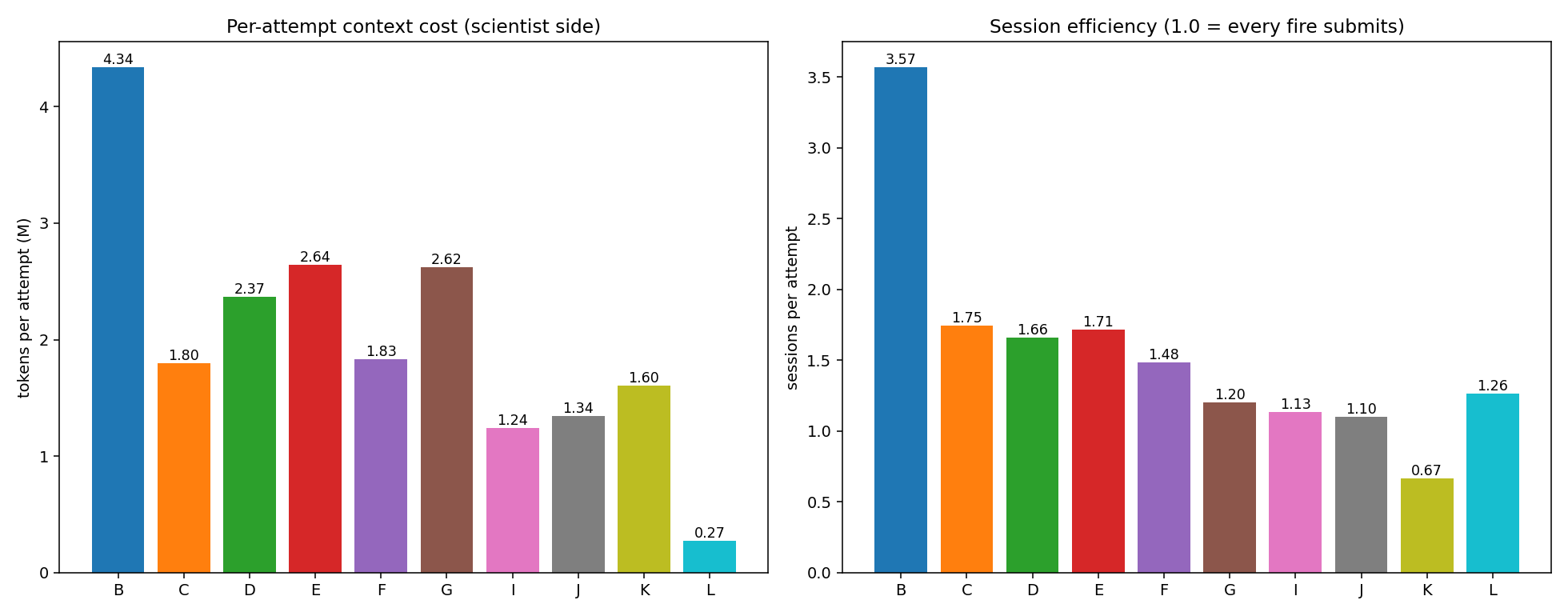}
\caption{Per-attempt economics (HCD window). Token attribution maps sessions to the HCD window via archive ledgers and is approximate for arms with many sessions per attempt.}
\label{fig:ablation_per_attempt}
\end{figure}

\paragraph{Ablation economics, reference block.}
Reference block for the ablation economics: Table~\ref{tab:value_ratio} reports the value-ratio composite (score at HCD per M-token per 0.01 gain); Figures~\ref{fig:ablation_gain_per_token} and \ref{fig:ablation_cost_per_point} show the underlying economics; Figure~\ref{fig:ablation_per_attempt} the per-attempt breakdown.

\begin{table}[H]
\centering
\caption{Value-ratio composite: score at HCD per M-token per 0.01 gain
(at-HCD mean divided by M-tokens-per-0.01-gain), with mean token spend per run
where recorded. The floor-prompt arm L wins on both axes: highest score and
${\sim}5{-}7\times$ cheaper iterations than F or J. L cleared the pre-data
sensitivity bound ($<3.08$ M/pt needed to beat J) by an order of magnitude.}
\label{tab:value_ratio}
\small
\begin{tabular}{lcccc}
\toprule
Arm & Score at HCD & M-tok / 0.01 gain & Value ratio & Tokens / run \\
\midrule
\textbf{L} & \textbf{0.890} & \textbf{0.355} & \textbf{2.507} & \textbf{6.9M} \\
J & 0.850 & 2.937 & 0.289 & 37.5M \\
F & 0.854 & 3.026 & 0.282 & 49.8M \\
C & 0.832 & 2.97  & 0.280 & --- \\
K & 0.811 & ---   & 0.248 & --- \\
I & 0.807 & 3.27  & 0.247 & --- \\
D & 0.843 & 3.76  & 0.224 & --- \\
E & 0.849 & 6.78  & 0.125 & --- \\
G & 0.846 & 6.84  & 0.124 & --- \\
B & 0.794 & 18.56 & 0.043 & 111M \\
\bottomrule
\end{tabular}
\end{table}

\begin{figure}[H]
\centering
\includegraphics[width=0.75\textwidth]{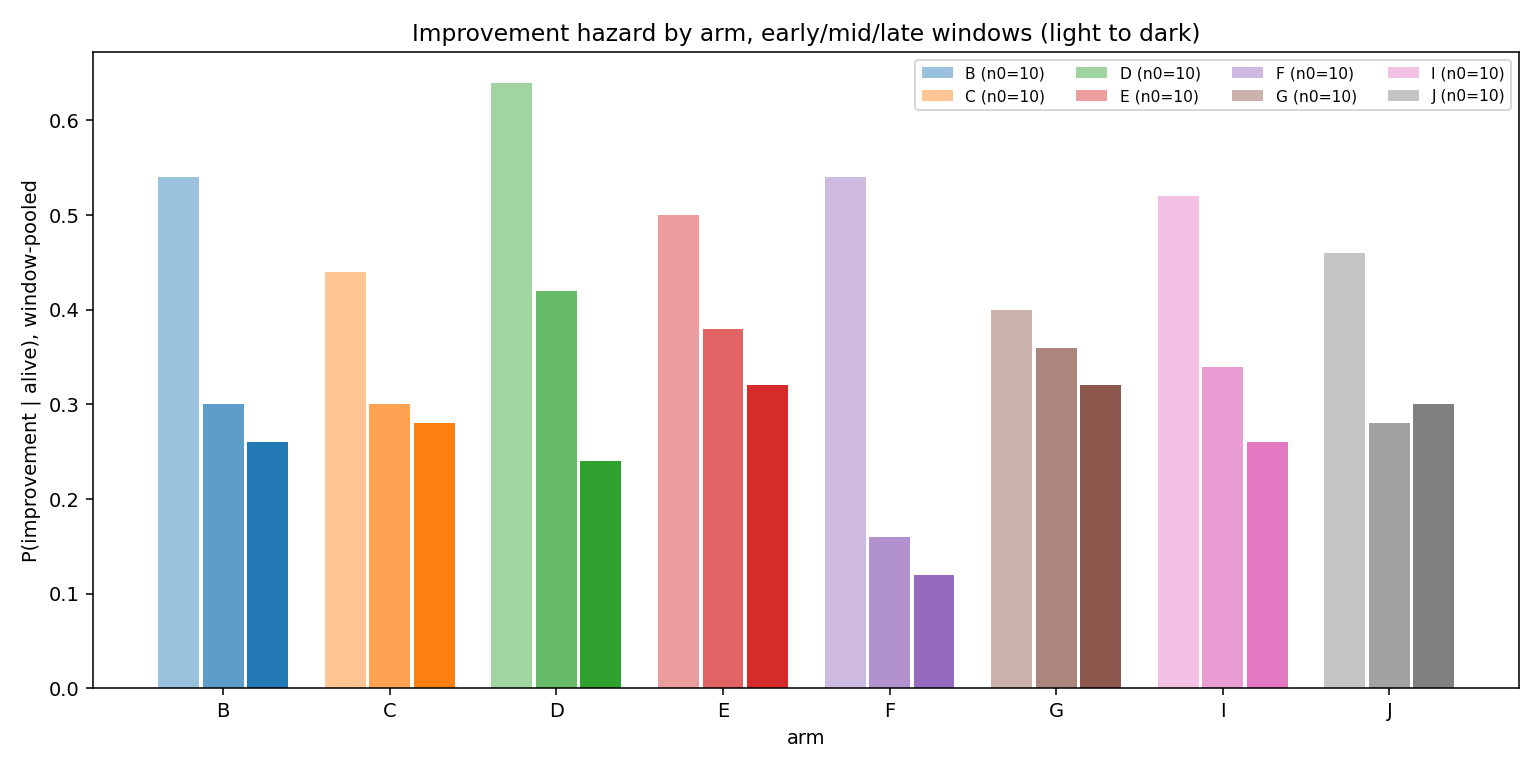}
\caption{Improvement hazard by arm: P(improve $\mid$ alive), pooled into early/mid/late windows (runs 1--5, 6--10, 11--15; light to dark). Every arm's conversion rate declines with iteration; no arm retains a distinct late-improvement channel.}
\label{fig:ablation_hazard}
\end{figure}

\begin{figure}[H]
\centering
\includegraphics[width=0.75\textwidth]{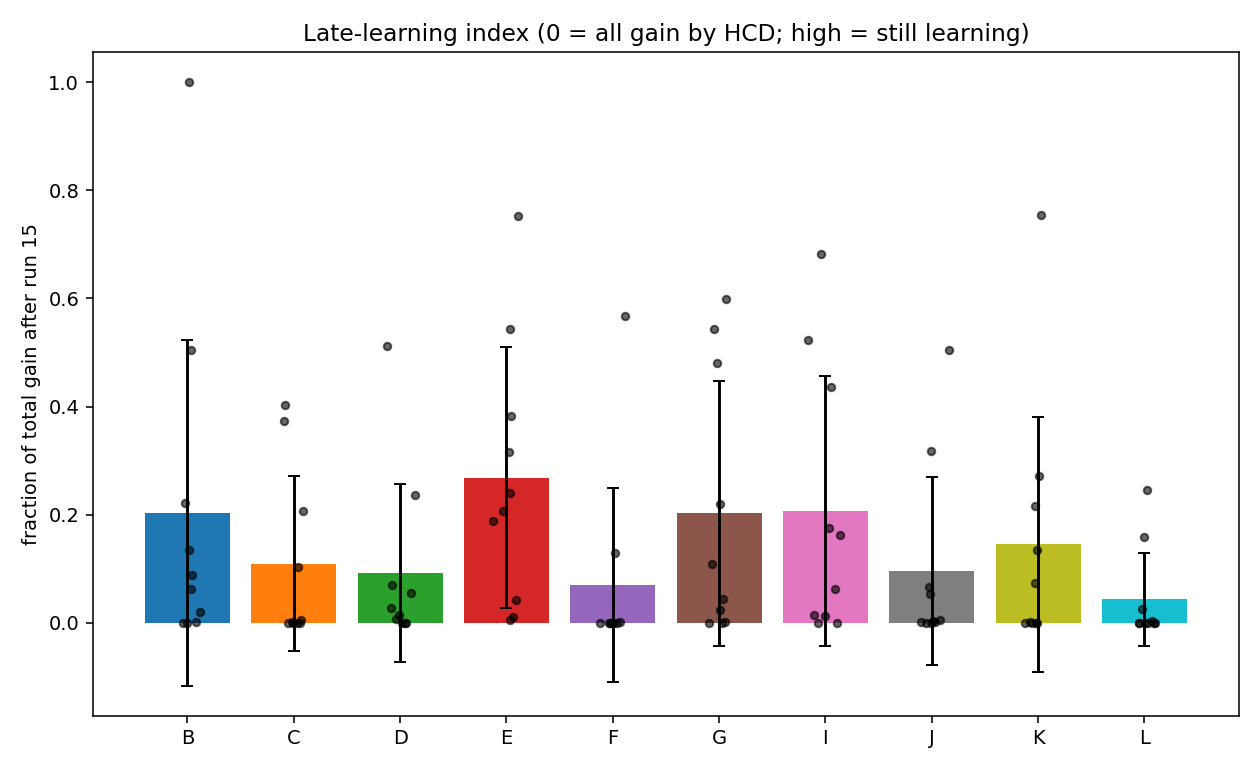}
\caption{Late-learning index: share of each arm's total gain earned after iteration 15.}
\label{fig:ablation_late}
\end{figure}

\paragraph{Improvement dynamics.}
Two further views complete the economics. Figure~\ref{fig:ablation_hazard} shows the improvement hazard --- the probability of improving at iteration $t$ conditional on still being run --- which shows improvement probability declining monotonically for every arm: attempts keep converting, but at a falling rate, and no arm retains a distinct late-iteration improvement channel. Figure~\ref{fig:ablation_late} shows the late-learning index (fraction of total gain occurring after the HCD window): arms with high late-learning indices were still earning score when the budget clock stopped, indicating headroom the 15-iteration cap did not let them spend.

B's 18.56M tokens per 0.01 gain is ${\sim}50\times$ L's 0.355M. Token forensics explain why: context-cache re-reads account for roughly $2.5\times$ the counted token total, while its output is ${\sim}5$M tokens/run ($4\%$ of spend) and explicit reasoning ${\sim}2.5\%$ --- the scaffold's cost is its context, and its outcome correlates almost entirely with its starting score ($r(\text{initial}, \text{top}) = +0.89$ for B).

\section*{Section G: Prior-Frontier Dynamics (Derivations and Sources)}

This appendix re-derives, from data on file, every number behind the Section~\ref{sec:discussion-mining}
boundary claims (``no process reliably exceeded what its prior contained'').

\begin{table}[H]
\centering
\small
\caption{Held-out frontier (Full TF Metric): process artifacts against
the prior's best cold draw. n=197 scored one-shot draws (pools H and
H$_G$). The baseline row is a reference, not a process output; the E[best-of-15] row is an Ablation Metric statistic.}
\begin{tabular}{llll}
\toprule
Artifact / process & Held-out & vs.\ prior max & Reading \\
\midrule
\midrule
\multicolumn{4}{l}{\emph{Baseline (not a process output):}} \\
Best cold draw (H\_4\_11; proxy 0.8023) & 0.8757 & --- & the prior's frontier (Full TF Metric) \\
Bandit bars: E[best-of-15] (H / H$_G$) & 0.8463 / 0.8455 & $-0.029$ & expected sampling yield (Ablation Metric) \\
B champion (loop, same package) & 0.8774 & $+0.0017$ & tie (draw noise) \\
L best artifact (floor-prompt loop) & 0.8740 & $-0.0017$ & parity \\
Pre-ablation champion (DALE, audited) & ${\approx}0.868 \pm 0.003$ & $-0.008$ & below \\
OpenEvolve / ShinkaEvolve / CodeEvolve (means) & 0.862 / 0.852--0.856 / 0.862 & below & band-level \\
\textbf{r5\_1} (Evolve framing rung) & \textbf{0.8819} (0.8964 STREME's Benchmark) & $+0.0062$ & the one clear pass \\
r5\_2 / r5\_3 (same rung, other draws) & 0.8052 / 0.7022 & far below & non-replication \\
\bottomrule
\end{tabular}
\end{table}

\paragraph{Derivations.}
(1)~\emph{Cold-draw frontier:} every H/H$_G$ one-shot artifact scored on the
53-TF benchmark ($n{=}197$); the maximum is draw H\_4\_11 at 0.8757 with
proxy eval 0.8023. Per-draw pairs (eval, benchmark) live in
\texttt{transfer-eval/results/slurm/H\_*\_P100.json}; recomputation scripts:
\texttt{goodhart\_selection.py}, \texttt{feedback\_effect\_quant.py} (repo).
(2)~\emph{Exchangeability test:} under the null that a process artifact is
exchangeable with pool draws, $P(\text{new} > \text{max of } 197) =
1/198 \approx 0.005$. r5\_1 passes ($+0.0062$); B's $+0.0017$ sits within
scoring noise of the maximum and is read as a tie.
(3)~\emph{Bandit bars:} E[best-of-15] via 5{,}000-draw Monte Carlo, seed 42
(registered nulls, Section~\ref{sec:ablation-nulls}).
(4)~\emph{DALE metric audit:} scoring-rule decomposition ($+0.0200$ scoring
rule, $+0.0047$ candidate count, $+0.0031$ implementation, $+0.0000$
language) re-anchors the packaged 0.891 (STREME's Benchmark) to
${\approx}0.868 \pm 0.003$ on the Full TF Metric; the audit text is recorded
in the repository history (commit \texttt{4927a53}\^{}).
(5)~\emph{Proxy maxima:} L 0.9990 and G\_5 1.0193 against pool maxima 0.869
(H) and 0.854 (H$_G$); G\_5's within-run held-out trajectory
$0.816 \to 0.844 \to 0.699$ (\S\ref{sec:ablation-genomic}).
(6)~\emph{L reliability:} 10/10 runs' top-run score at or above the bar;
top-run mean 0.8906 ($+0.044$; bootstrap 95\% CI $[0.846, 0.938]$); per-run
endpoints from \texttt{hcd\_run\_level.csv} and the run histories.

\paragraph{Guards.} L has no same-package pool (floor prompt never drew
one): its comparisons are cross-package. The Evolve family's true prior ---
the CodeEvolve engine's proposal distribution --- was never sampled cold;
r5\_1 is compared against the TLS pools' frontier, cross-family. Both
caveats make the boundary claim conservative: relaxing either would only
widen the prior's frontier.

\end{document}